\documentclass[aps,pra, twocolumn, superscriptaddress, floatfix]{revtex4-2}

\usepackage[margin=1in]{geometry}

\usepackage[T1]{fontenc}
\usepackage{bbm}
\usepackage{hyperref}
\hypersetup{%
	breaklinks=true,
	colorlinks,
	linkcolor={cyan},
	citecolor={cyan},
	urlcolor={cyan},
	pdfauthor = {Aniruddha Bapat, Stephen P. Jordan},
	pdftitle ={Approximate optimization of MAXCUT with a local spin algorithm},
	pdfpagemode=UseOutlines,
	pdfborder={0 0 0}
}

\usepackage{titlesec}
\titleformat{\section}[block]{\bfseries}{\thesection}{5pt}{\filcenter\MakeUppercase}{}
\titlespacing*{\section}{0pt}{*4}{*2}
\titleformat{\subsection}[runin]{\bfseries}{\thesubsection}{5pt}{\scshape}{}
\titlespacing*{\subsection}{0pt}{*1}{*2}

\usepackage{graphicx, subcaption}
\graphicspath{{plots/}}

\usepackage{amssymb,amsmath,amsthm}

\usepackage[nameinlink,capitalise]{cleveref}
\usepackage{booktabs, multirow, float}

\usepackage[toc,page, title,titletoc]{appendix}

\usepackage{animacros}

\begin{document}

\title{Estimating gate complexities for the site-by-site preparation of fermionic vacua}

\author{Troy Sewell}
\email{tjsewell@umd.edu}
\affiliation{Joint Center for Quantum Information and Computer Science, University of Maryland, College Park, Maryland 20742, USA}
\author{Aniruddha Bapat}
\email{ani@lbl.gov}
\affiliation{Joint Center for Quantum Information and Computer Science, University of Maryland, College Park, Maryland 20742, USA}
\affiliation{Lawrence Berkeley National Laboratory, Berkeley, CA 94720}
\author{Stephen P. Jordan}
\email{stephen.jordan@microsoft.com}
\affiliation{Microsoft, Redmond, WA 98052, USA}
\affiliation{University of Maryland, College Park, MD 20742, USA}


\date{\today}

\begin{abstract}

An important aspect of quantum simulation is the preparation of physically interesting states on a quantum computer, and this task can often be costly or challenging to implement. A digital, ``site-by-site'' scheme of state preparation was introduced in \cite{moosavian2019} as a way to prepare the vacuum state of certain fermionic field theory Hamiltonians with a mass gap. More generally, this algorithm may be used to prepare ground states of Hamiltonians by adding one site at a time as long as successive intermediate ground states share a non-zero overlap and the Hamiltonian has a non-vanishing spectral gap at finite lattice size. In this paper, we study the ground state overlap as a function of the number of sites for a range of quadratic fermionic Hamiltonians. Using analytical formulas known for free fermions, we are able to explore the large-$N$ behavior and draw conclusions about the state overlap. For all models studied, we find that the overlap remains large (\emph{e.g.} $> 0.1$) up to large lattice sizes ($N=64,72$) except near quantum phase transitions or in the presence of gapless edge modes. For one-dimensional systems, we further find that two $N/2$-site ground states also share a large overlap with the $N$-site ground state everywhere except a region near the phase boundary. Based on these numerical results, we additionally propose a recursive alternative to the site-by-site state preparation algorithm.

\end{abstract}

\maketitle

 The preparation of non-trivial quantum states is an important subroutine in quantum simulation. The canonical example is the (static) problem of preparing a quantum state that approximates the ground state of a Hamiltonian of interest, such as the electronic structure Hamiltonian in quantum chemistry or the Hamiltonian of a quantum field theory. In many cases, state preparation is also a necessary first step in the simulation of quantum dynamics, where one initializes a state that is expected to evolve in time with interesting dynamics. The algorithm proposed in~\cite{jordan2014} for simulating scattering in a scalar quantum field theory first involves a preparation of the vacuum of the interacting field theory, a subroutine that dominates the runtime cost of the full algorithm. It is therefore a well-motivated problem to design schemes for preparing the ground state of a known Hamiltonian. While such an algorithm cannot hope to be efficient in general due to the QMA-hardness of approximating the ground states of $k$-local Hamiltonians~\cite{kitaev2002}, it is nonetheless useful to understand when an efficient approximation is possible.    


Classical methods for simulating real-time dynamics are often limited by their inability to accurately simulate interference effects (also known as the ``sign problem''). The quantum approach is more natural: for a properly discretized field theory, one can map the local degrees of freedom directly to qubits of the quantum device, then prepare an initial state that corresponds to the state of interest (say a scattering state), and engineer interactions mimicking the true form of interaction seen in nature (analog) or provide a quantum circuit description of the time evolution (digital).  

For theories in 1+1D spacetime, a fast state initialization algorithm was proposed~\cite{moosavian2018} using ideas from tensor networks. A similarly motivated proposal used projected entangled pair states (PEPS) to prepare the ground state of a lattice in two spatial dimensions~\cite{schwarz2012}. However, both proposals rely on knowing the tensor network description of the ground state and carry a (possibly large) classical computational overhead.  

Recently in \cite{moosavian2019}, an algorithm was introduced for the preparation of vacuum states of lattice Hamiltonians digitally on a quantum computer. Starting from the ground state $\ket{g_{N_0}}$ of the target Hamiltonian on a constant number of sites $N_0$, the algorithm prepares successively larger ground states by adding one site at a time to the boundary of the lattice, until the number of sites equals the target size, $N$. Each site addition step involves a fixed-point Grover search~\cite{yoder2014} routine whose oracle calls invoke quantum phase estimation on the underlying ground states. The quantum phase estimation subroutine, in turn, uses time evolution under the target Hamiltonian. Therefore this ``site by site'' state preparation algorithm requires the ability to simulate the target Hamiltonian but is otherwise oblivious to the details of the ground state.  
Two additional conditions are assumed, for each $n\in\{N_0,\ldots, N-1,N\}$:
\begin{enumerate}
    \item The spectral gap $m_n$, or the difference between the energies of the first excited state and ground state, is non-zero: $m_n > 0$.
    \item The successive ground state overlaps $\eta_n:= |\paren{\bra{g_n}\otimes\bra{q}}\ket{g_{n+1}}|$ are non-zero. 
\end{enumerate}
The second condition assumes the freedom to choose the state $\ket{q}$ on the added site(s) to maximize overlap.

On Hamiltonians that are geometrically local with each term bounded in spectral norm, the gate complexity of the site-by-site algorithm is $O\paren{\frac{n}{m_n\eta_n}\log (1/\epsilon)}$ for the iteration $\ket{g_n}\mapsto \ket{g_{n+1}}$. Success therefore crucially depends on how $m_n$ and $\eta_n$ scale with the number of sites $n$. In particular, for a Hamiltonian with an exponentially closing gap, the algorithm fails to have a polynomial-time guarantee. 

On the other hand, when the spectral gap is constant, as is the case for massive field theories, we have $\lim_{n\rightarrow\infty}m_n = O(1)$, giving a trivial contribution to the runtime. Reference~\cite{moosavian2019} uses tensor network techniques to study the scaling of overlap for the massive Gross-Neveu model. The numerical calculations presented there on lattices up to $N=50$ suggest that, for various settings of the coupling and bare mass, the overlap converges to a constant in the limit of large $N$. Similar results are obtained for the AKLT model. More generally, for models of interacting field theories, it is not known how the state overlap behaves as a function of system size and other model parameters. 

Non-interacting models are useful for studying the behavior of physical properties over a broad range of system parameters, since they exhibit rich physics but are often computationally tractable. In the case of free fermions, the Hamiltonian, which is quadratic in the mode creation and annihilation operators, is diagonalizable in time $O(n^3)$ where $n$ is the number of sites. The ground states of quadratic fermionic Hamiltonians are Gaussian, i.e. fully characterized by the expectation values of all degree-two monomials in the mode operators. As shown in~\cite{bravyi2017}, the magnitude of the inner product of any two fermionic Gaussian states can be exactly computed classically in time $O(n^3)$, while the gap is obtained directly as the smallest mode energy of the diagonalized Hamiltonian. Here, we use this feature to study the dependence of the spectral gap and the overlap for a range of free fermionic Hamiltonians, and provide concrete estimates for the runtime of the site-by-site algorithm on these models. 

Although free fermion models are exactly solvable and therefore do not pose a computational challenge requiring quantum computers, they provide a tractable test case to investigate the general behavior of the ground state overlaps determining the performance of site-by-site quantum state preparation. As discussed in sections II-V, extrapolation to weakly coupled fermionic systems is possible through rigorous methods, and conjectures about behavior beyond weak coupling can be generated from the patterns observable in numerical data from the free case.

The paper is organized as follows. In \cref{sec:covariance} we describe the free fermion formalism. Our results follow in~\cref{sec:numerics}, in which we explore the phase diagrams as well as the large-$N$ behavior of the spectral gap and overlap for a range of models. In~\cref{sec:recursive} we introduce a possible improvement to the site-by-site algorithm for one-dimensional systems, and we conclude with~\cref{sec:conclusion}. In appendix \ref{app_schmidt} we then discuss some connections of the state overlap with the Schmidt spectrum. 

\section{Free fermions and the covariance matrix formalism}
\label{sec:covariance}
We start with a short overview of the covariance matrix formalism for (fermionic) Gaussian states. A system of  non-interacting, spinless fermions on $N$ sites can be modeled by a Hamiltonian which is quadratic in the mode creation and annihilation operators $c_j\dg, c_j$ on site $j=1,\ldots N$, is given by
\begin{equation}
\label{eq:HDirac}
    H = \suml{i,j}{}{A_{ij} c_i\dg c_j} + \frac{1}{2}\suml{i,j}{}{(B_{ij}c_i\dg c_j\dg + h.c.)}\, .
\end{equation}
The mode operators satisfy  anti-commutation relations $\{c_i,c_j\} = 0, \{c_i,c_j\dg\} = \delta_{ij}$, which allows us to restrict the form of the coupling matrices $A, B$. Namely, we assume that $A_{ij} = A_{ji}^*$ (Hermitian), and $B_{ij}=-B_{ji}$ (anti-symmetric). The Hamiltonian can then be written more compactly in a bilinear form, 
\begin{equation}
\label{eq:HDiracCoupling}
       H = \frac{1}{2}\begin{bmatrix}\vec{c}\dg& \vec{c}\end{bmatrix}\begin{bmatrix}A & B\\-B^* & -A^*\end{bmatrix}\begin{bmatrix}\vec{c}\\ \vec{c}\dg\end{bmatrix}\, ,
\end{equation}
where we omit an overall additive constant $\frac{1}{2}\text{Tr}(A)$. The same Hamiltonian can also be expressed in terms of Majorana operators which we define for a given site $j$ as
\begin{equation}
\label{eq:MajoranaDefinition}
    \gamma_j = \frac{(c_j+c_j\dg)}{\sqrt{2}},\     \gamma_j' \equiv \gamma_{j+N} =  \frac{(c_j-c_j\dg)}{\sqrt{2}i}\, , 
\end{equation}
Like their Dirac counterparts, the Majorana operators satisfy canonical anti-commutation relations $\curly{\gamma_i,\gamma_j} = \delta_{ij}$. In this form, the Hamiltonian can be expressed as
\begin{equation}
\label{eq:HMajoranaCoupling}
    H = \frac{i}{2}\begin{bmatrix}\vec{\gamma}& \vec{\gamma}'\end{bmatrix}\begin{bmatrix}\text{Im}(A-B) & \text{Re}(A-B)\\\text{Re}(A+B) & \text{Im}(A+B)\end{bmatrix}\begin{bmatrix}\vec{\gamma}\\ \vec{\gamma}'\end{bmatrix}\, .
\end{equation}
It can be verified that the above coupling matrix is traceless and anti-Hermitian. 

The ground states (and all excited states as well for fermions) of a quadratic Hamiltonian as given in~\cref{eq:HMajoranaCoupling} are Gaussian, and therefore are fully characterized by all linear and quadratic expectation values $\langle\gamma_i \rangle$, $\langle\gamma_i \gamma_j\rangle$. Parity conservation (which follows from $H$ having a quadratic form) implies that the first moments $\langle\gamma_i\rangle$ vanish. The expectation values of all quadratic products of the Majorana modes can be arranged in the form of the \emph{covariance matrix} given by
\begin{equation}
    \Gamma_{jk} = i\langle[\gamma_j,\gamma_k]\rangle \equiv i\langle\gamma_j\gamma_k - \gamma_k\gamma_j\rangle\, .
\end{equation}
$\Gamma$ is real, Hermitian, and anti-symmetric by construction, and fully characterizes the state. It can also be shown that the eigenvalues of $\Gamma$ are $\pm i$.  

Consider an arbitrary basis change that transforms $\curly{\gamma_j}$ to $\curly{\lambda_j}$ via the linear map $\lambda_j = O_{jk}\gamma_k$. Since the $\lambda_j$ must satisfy anti-commutation relations, it follows that 
\begin{align}
    \delta_{ij} &= \curly{\lambda_i, \lambda_j}\\ 
        &= \suml{k,l}{}{O_{ik}O_{jl}\curly{\gamma_k, \gamma_l}} = \suml{k,l}{}{O_{ik}O_{jl}\delta_{kl}} \\
        &= \suml{k}{}{O_{ik}O_{kj}^T} = [OO^T]_{ij}\, ,
\end{align}
which implies that $O$ is an element of the orthogonal group $SO(2n)$ (and conversely, any orthogonal matrix corresponds to a valid basis transformation). The covariance matrix transforms under $O$ as
\begin{align}
    \Lambda_{ij} &= i\langle[\lambda_i, \lambda_j]\rangle\\ 
        &= i\suml{k,l}{}{O_{ik}O_{jl}[\gamma_k, \gamma_l]} = i\suml{k,l}{}{O_{ik}O_{jl}\Gamma_{kl}} \\
        &= i\suml{k, l}{}{O_{ik}\Gamma_{kl} O_{lj}^T} = [O\Gamma O^T]_{ij}\, .
\end{align}
In other words, $\Lambda = O\Gamma O^T$. Any real antisymmetric matrix $M$ can be expressed in a canonical form $M = P^T\Omega_{M} P$ for some orthogonal matrix $P$, where $\Omega_{M}$ is given by 
\begin{equation}
    \Omega_M = \begin{bmatrix}O & E_M \\ -E_M & O\end{bmatrix}\, ,
\end{equation}
where $O$ is the null matrix. The matrix $E_M$ is a diagonal matrix containing the eigenvalues of $M$ times $i$. For the covariance matrix $\Gamma$, we can write $\Gamma = P^T\Omega P$, where $\Omega = \begin{bmatrix}O & I \\ -I & O\end{bmatrix}$ is known as the \emph{symplectic form}.
The modes corresponding to this transformation, $\lambda_j = P_{jk}\gamma_k$, are the eigenmodes of the system. 

Revisiting the Hamiltonian in~\cref{eq:HMajoranaCoupling}, we can write the (real antisymmetric) coupling matrix as $P^T \begin{bmatrix}O & E \\ -E & O\end{bmatrix} P$ where $P$ is diagonalizing transformation. This brings $H$ into decoupled form 
\begin{equation}
    H = \frac{i}{2}\suml{}{}{E_{j}\lambda_j\lambda_j'}\, ,
\end{equation}
where $\curly{E_i}$ are the energies of the eigenmodes of $H$. The difference between the energies of the vacuum (no modes occupied) and the first excited state (lowest energy mode occupied), or the spectral gap, is simply given by the lowest mode energy, $m := \min_{j}{E_j}$.

The transformation $P$ that brings the ground state covariance matrix $\Gamma$ into canonical form also diagonalizes $H$. This provides a way to compute $\Gamma$ from the coupling matrix of the Hamiltonian. Namely, one finds the matrix $P$ via the diagonalization of the coupling matrix, and then computes $\Gamma = P^T\Omega P$.   

The covariance matrix organizes information about a Gaussian state in a useful form. Given two pure Gaussian states $\ket{\psi_1}, \ket{\psi_2}$ and corresponding covariance matrices $\Gamma_1, \Gamma_2$, the magnitude of overlap between them, $|\braket{\psi_1}{\psi_2}|$, has a particularly simple form
\begin{equation}
   |\braket{\psi_1}{\psi_2}| = \det \left(\frac{\Gamma_1+\Gamma_2}{2}\right)^{1/4}\, .
\end{equation}
(For a proof, see, e.g.,~\cite{bravyi2017}.)

\section{Numerical Results}
\label{sec:numerics}
\subsection{Kitaev Chain}
We begin with the Kitaev chain, a model of spinless fermions on a lattice in one spatial dimension, growing the open boundary condition ground state site-by-site. The Hamiltonian has single-site terms with coefficient $-\mu$ and nearest-neighbor coupling terms with coefficients $t$ for hopping and $\Delta$ for squeezing. 

\begin{equation}
    H_{nn} = -\mu \sum_{j=1}^N c_j^\dagger c_j -  \sum_{j=1}^{N-1}  \left ( t c_j^\dagger c_{j+1} + \Delta c^\dagger_j c^\dagger_{j+1}  \right ) + h.c.
\end{equation}

\begin{figure}[tb]
\centering
  \includegraphics[width=0.95\linewidth]{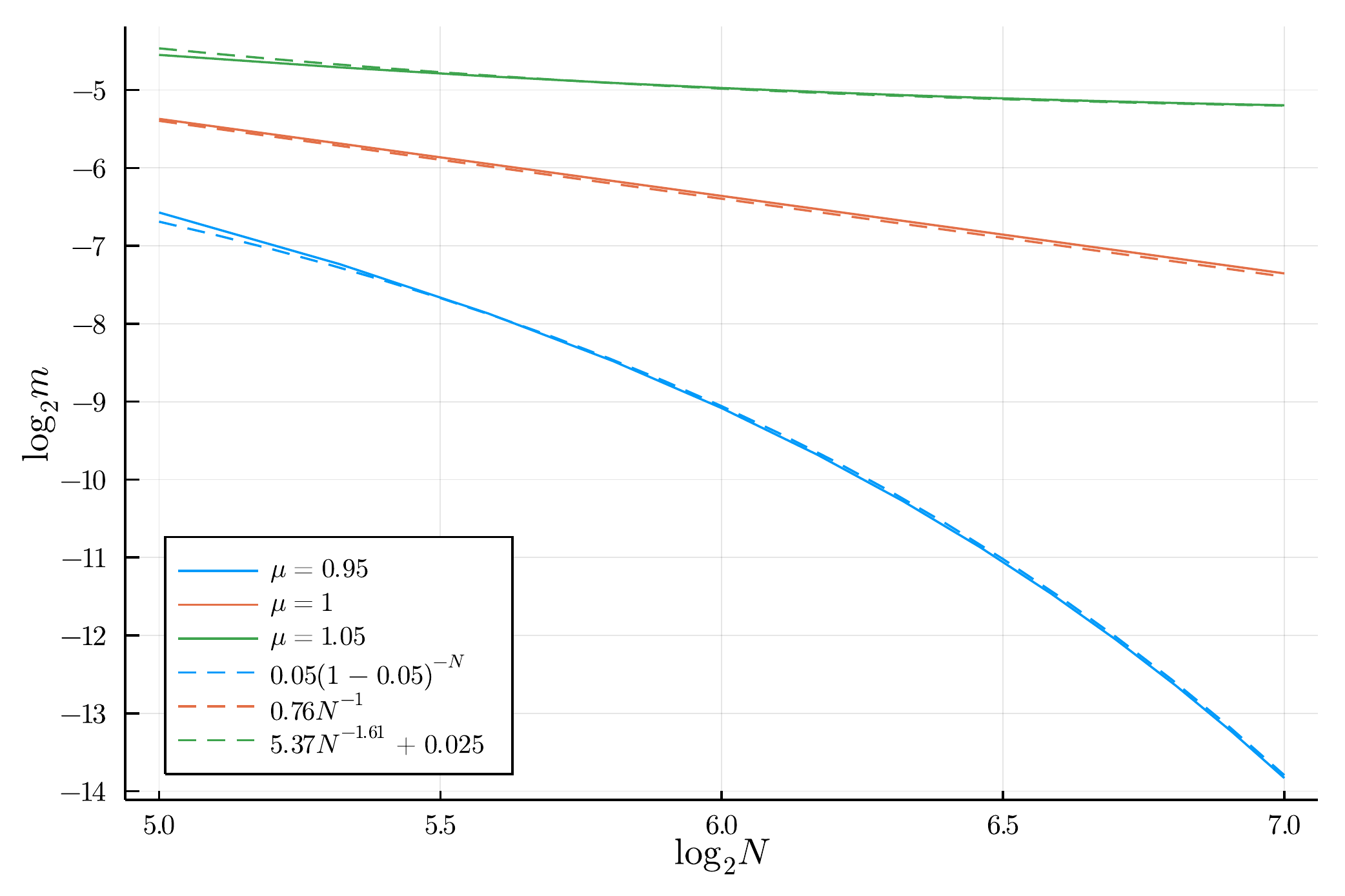}
  \includegraphics[width=0.95\linewidth]{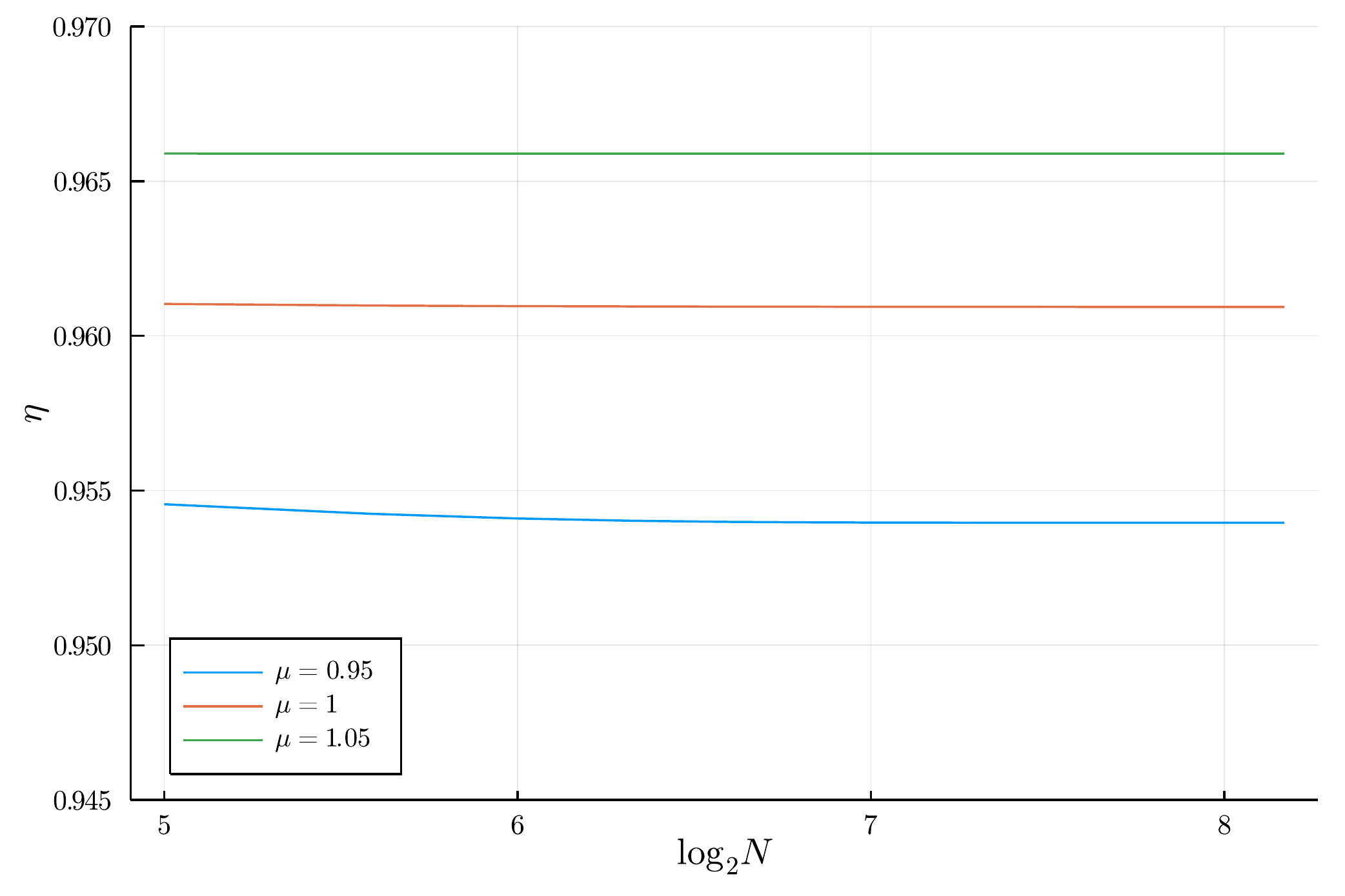}
\caption{Scaling of the energy gap(top) and overlap(bottom) with system size in the Kitaev chain for three values of $\mu$ along the line $\Delta =1$ near the critical point at $\mu=1$. The three points chosen are representative of the two phases on either side of the phase transition at $\mu=1$ and the critical model at the phase transition itself. Best fit curves are drawn in dashed lines on the log-log plot for the energy gap, showing behavior consistent with asymptotically constant, polynomial, and exponential scaling. The overlaps for these regions show convergence to large constant values with large system size.  }
\label{fig:gapN1d}
\end{figure}

Fixing $t=1$, the phase diagram has critical lines at $|\mu| = 1$, and $\Delta = 0, |\mu|<1$. These critical lines feature an inverse polynomially closing energy gap for finite system sizes. The regions with $|\mu| > 1$ are the topologically trivial gapped phases, while the regions with $|\mu| < 1$ are the topological phases. The trivial phase has an asymptotically constant gap, while the topological phases with open boundary conditions support Majorana zero modes with a gap closing exponentially in the system size. The scaling of the gap with system size for these three phases is shown in Fig. \ref{fig:gapN1d}.

\begin{figure}[tb]
\centering
  \includegraphics[width=0.95\linewidth]{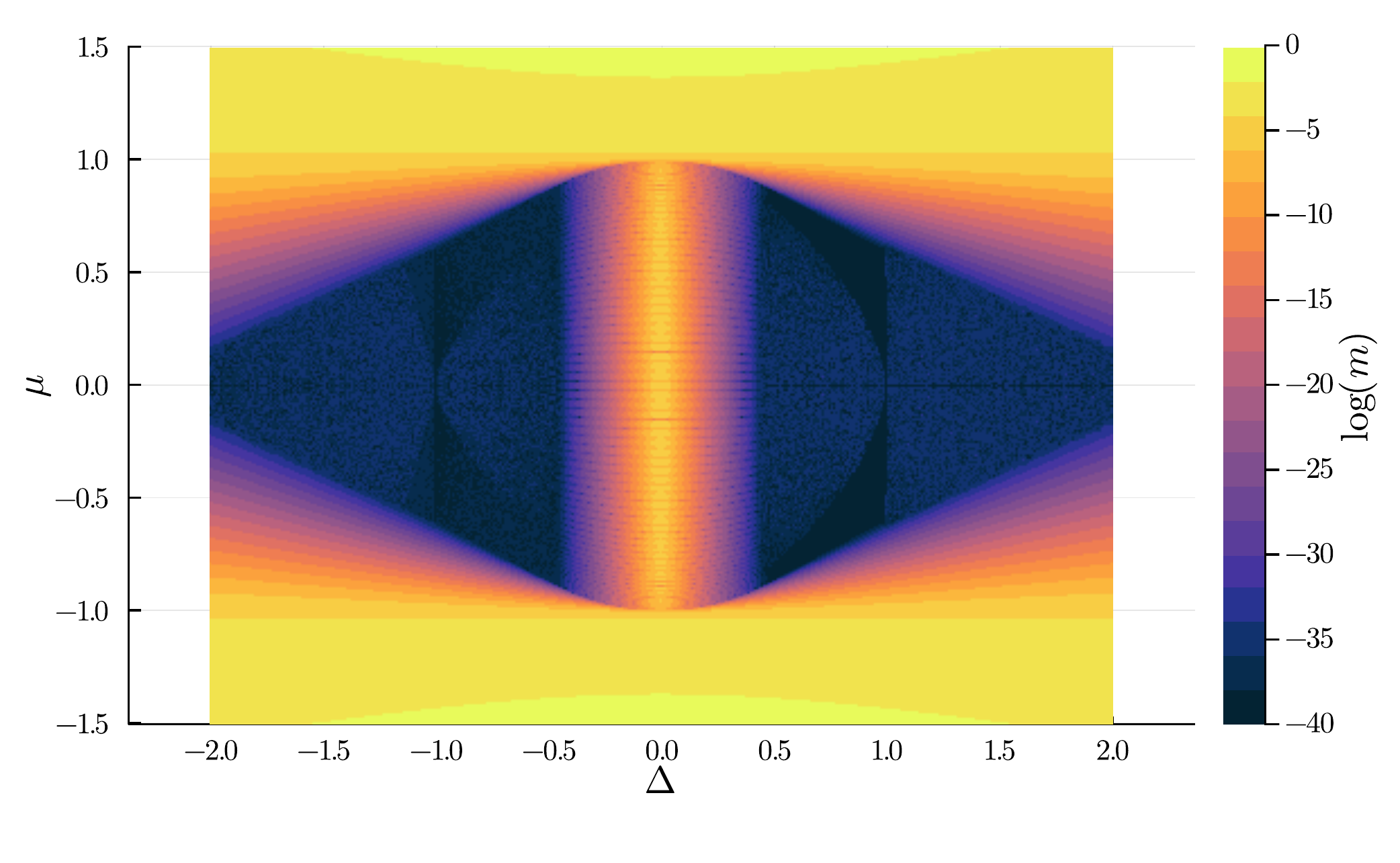}
\caption{Energy gap $m$ for 1D spinless fermions, Kitaev chain open boundary conditions, $N=72$, with logarithmic color scale taken base $e$.}
\label{fig:1dgap}
\end{figure}

\begin{figure}[tb]
\centering
  \includegraphics[width=0.95\linewidth]{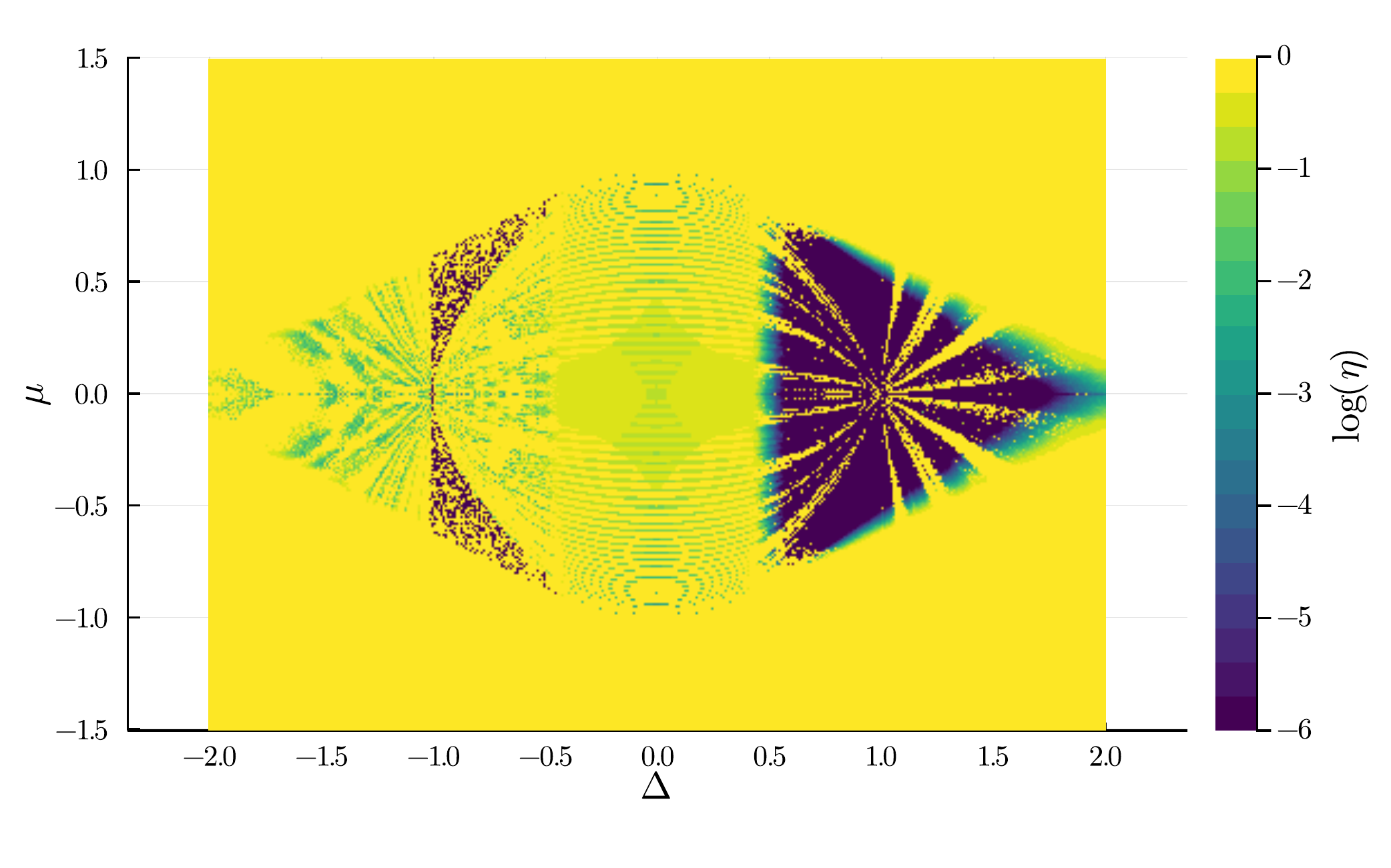}
\caption{Ground state overlap $\eta$ when adding a single site for spinless fermion Kitaev chain with open boundary conditions, $N=72$, with logarithmic color scale taken base $e$.}
\label{fig:1dover}
\end{figure}

The finite system size scaling for the ground state overlaps at these same points in the phase diagram are also shown in Fig. \ref{fig:gapN1d}. Despite the energy gap closing with system size for the critical point and point in the topological region, the ground state overlaps appear to converge to large constant values for each of these points at large system sizes. This can be seen by comparing Fig. \ref{fig:1dgap} and Fig. \ref{fig:1dover}, where the critical line at $\mu =1$ marks where the gap begins to close in Fig. \ref{fig:1dgap}, while nothing special appears to happen along this line in Fig. \ref{fig:1dover}. 

The overlap of the target ground state with the ground state of one smaller system, plus one additional fermion, is relatively large everywhere except the critical line along $\Delta=0$ and areas of the disordered phase contained in the topological region where $|\mu| < 1$. 
Considering that we can choose the single site state optimally, this amounts to the fidelity of the $N-1$ site ground state with the subystem of the $N$ site ground state. We expect these to have high overlap where there is small correlation length, due to edge effects having negligible effect on most of the state. However the overlaps appear to be large elsewhere, even along the critical lines.

The fidelity for mixed Gaussian states can be computed using a more complicated formula in terms of their respective covariance matrices \cite{Recovery19}. For numerical stability we used very low temperature states as the ground states. In the topological phases where the gap is closing exponentially, there effectively becomes a degenerate ground space. Working with pure states rather than very low temperature states as in this numerical plot, this overlap may be achievable by choosing the correct state in the effectively degenerate ground state space. However, since the ground space is effectively degenerate the algorithm may fail anyway due to the exponentially small gap.
Another possibility could be to work in the periodic fermion chain where the topological phases are gapped. 

\subsection{2D nearest-neighbor}
In two spatial dimensions we consider an analogous nearest neighbor coupled spinless fermion Hamiltonian with open boundary conditions. The state can be built up site-by-site around the boundary to successively larger square lattices with $\ell$ fermions along each dimension. 

\begin{equation}
\begin{split}
    H_{2D} = &-\mu \sum_{i,j=1}^\ell c_{i,j}^\dagger c_{i,j} -  \sum_{i,j=1}^{\ell-1}  \left( t( c_{i,j}^\dagger c_{i,j+1} + c_{i,j}^\dagger c_{i+1,j}) \right.\\
    &+ \left. \Delta ( c^\dagger_{i,j} c^\dagger_{i,j+1}  + c^\dagger_{i,j} c^\dagger_{i+1,j})\right) + h.c.
\end{split}
\end{equation}

\begin{figure}[tb]
\centering
  \includegraphics[width=0.95\linewidth]{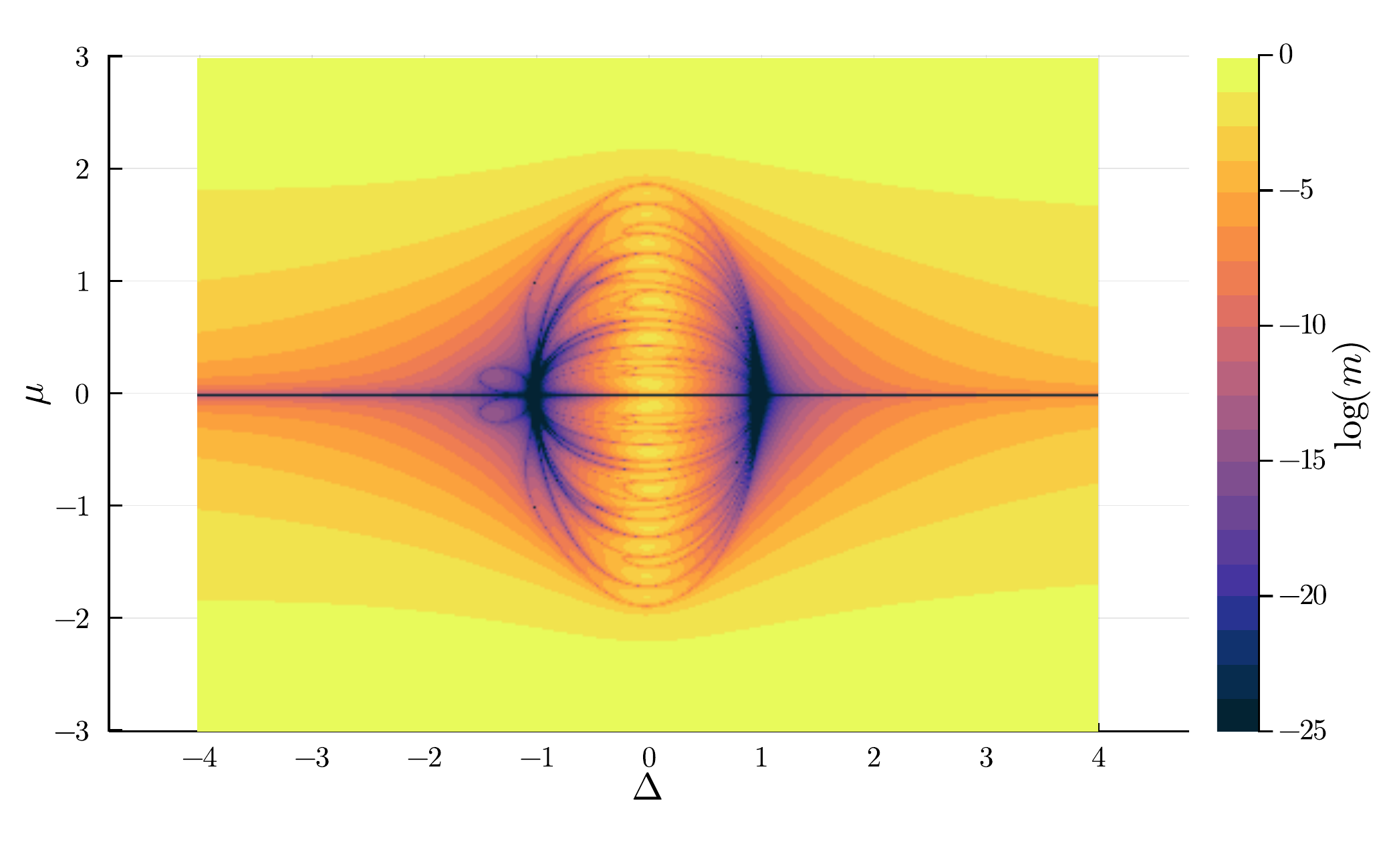}
\caption{Energy gap $m$ for 2D spinless fermions, square lattice with side length $\ell = 8$, with logarithmic color scale taken base $e$.}
\label{fig:2dgap}
\end{figure}

\begin{figure}[tb]
\centering
   \includegraphics[width=0.95\linewidth]{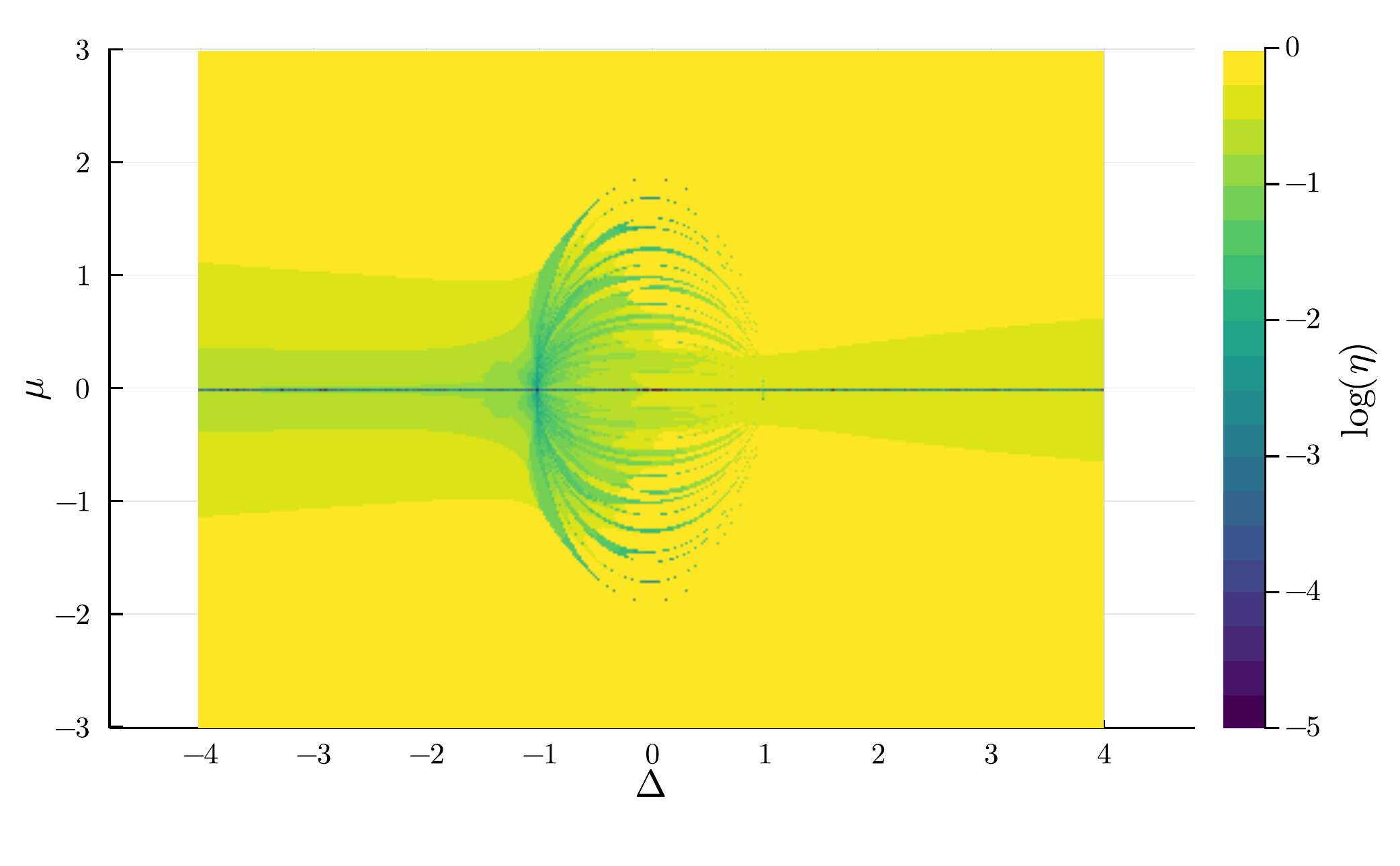}
\caption{Ground state overlap $\eta$ when adding a single site on the corner to complete square 2D lattice of spinless fermions with side length $\ell = 8$. Color scale is logarithmic taken base $e$.}
\label{fig:2dover}
\end{figure}

\subsection{Globally coupled}
We also consider a globally coupled model with analogous quadratic terms. This model lacks spatial locality, however there is still an ordering of the fermions which determines the sign of the various coupling terms. 
\begin{equation}
    H_0 =-\mu \sum_{j=1}^N c_j^\dagger c_j - \sum_{j=2}^{N} \sum_{l=1}^{j-1} \left (  t c_j^\dagger c_{j+l} + \Delta c^\dagger_j c^\dagger_{j+l}  \right ) + h.c.
\end{equation}

\begin{figure}[tb]
\centering
  \includegraphics[width=0.95\linewidth]{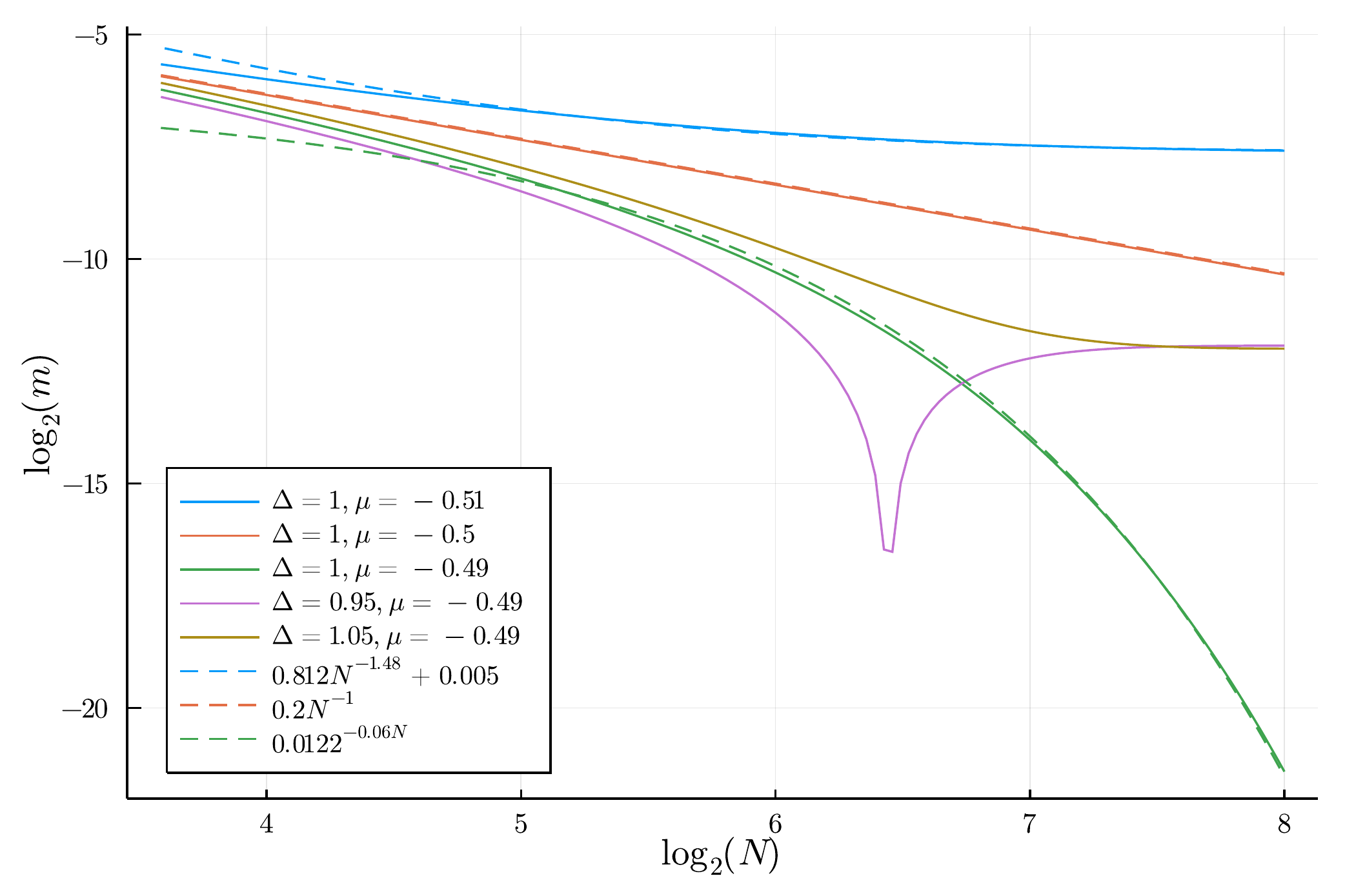}
   \includegraphics[width=0.95\linewidth]{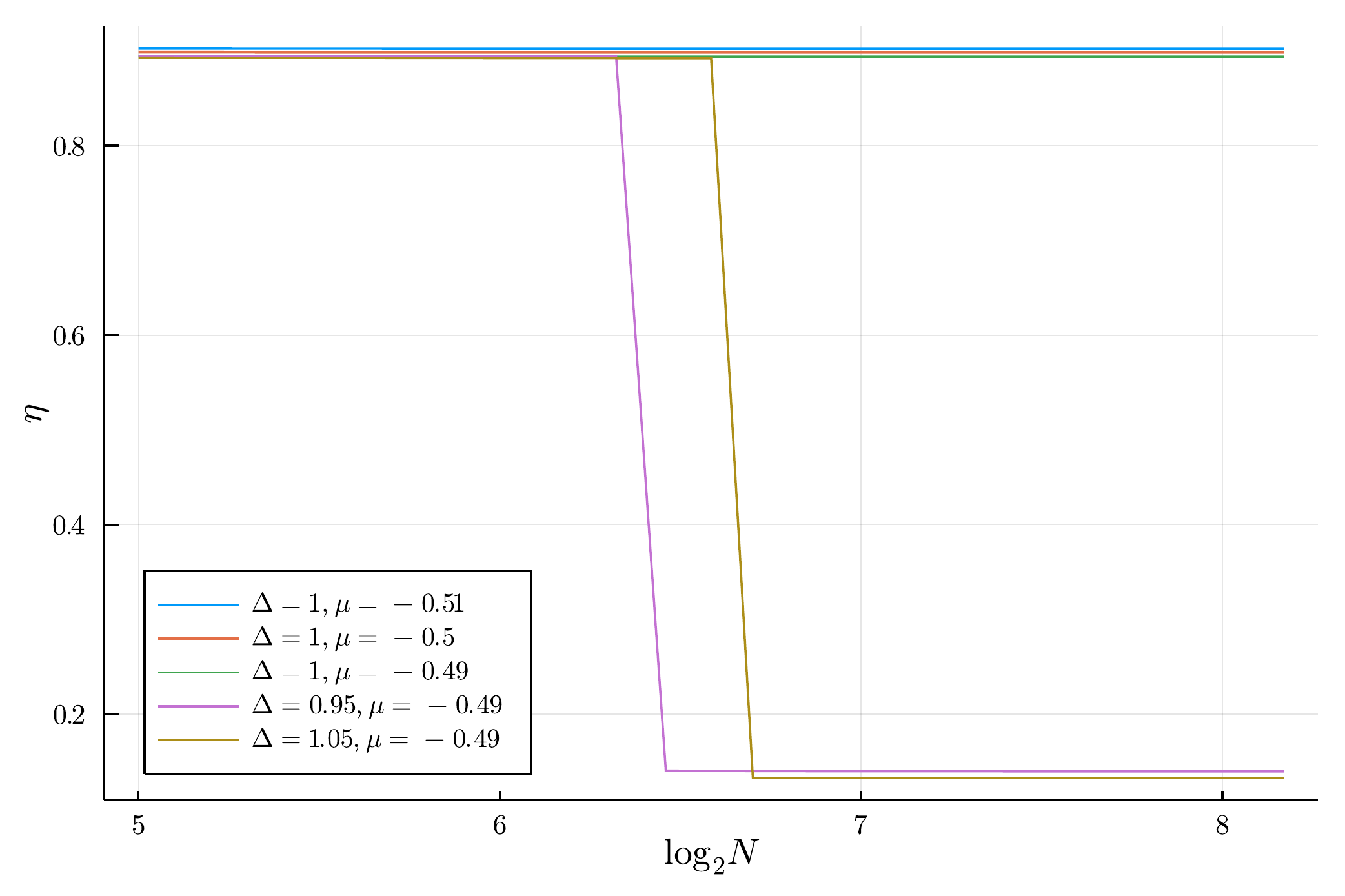}
\caption{Scaling of the energy gap(top) and ground state overlap(bottom) with system size for several points in the phase diagram of the globally coupled model around $\mu = -1/2, \Delta = 1$. These points are chosen to be representative of the asymptotically constant gap for the several extensive regions, the polynomially closing gap along the line $\mu = -1/2$ and an exponentially closing gap along the lines $\Delta = \pm 1$ with $\mu > -1/2$. Best fit curves are drawn for the the points along $\Delta = 1$ in dashed lines for the energy gap scaling, showing asymptotically constant, polynomial, and exponential behavior. The overlaps for most of these points are large and constant.}
\label{fig:gapNG}
\end{figure}

\begin{figure}[tb]
\centering
  \includegraphics[width=0.95\linewidth]{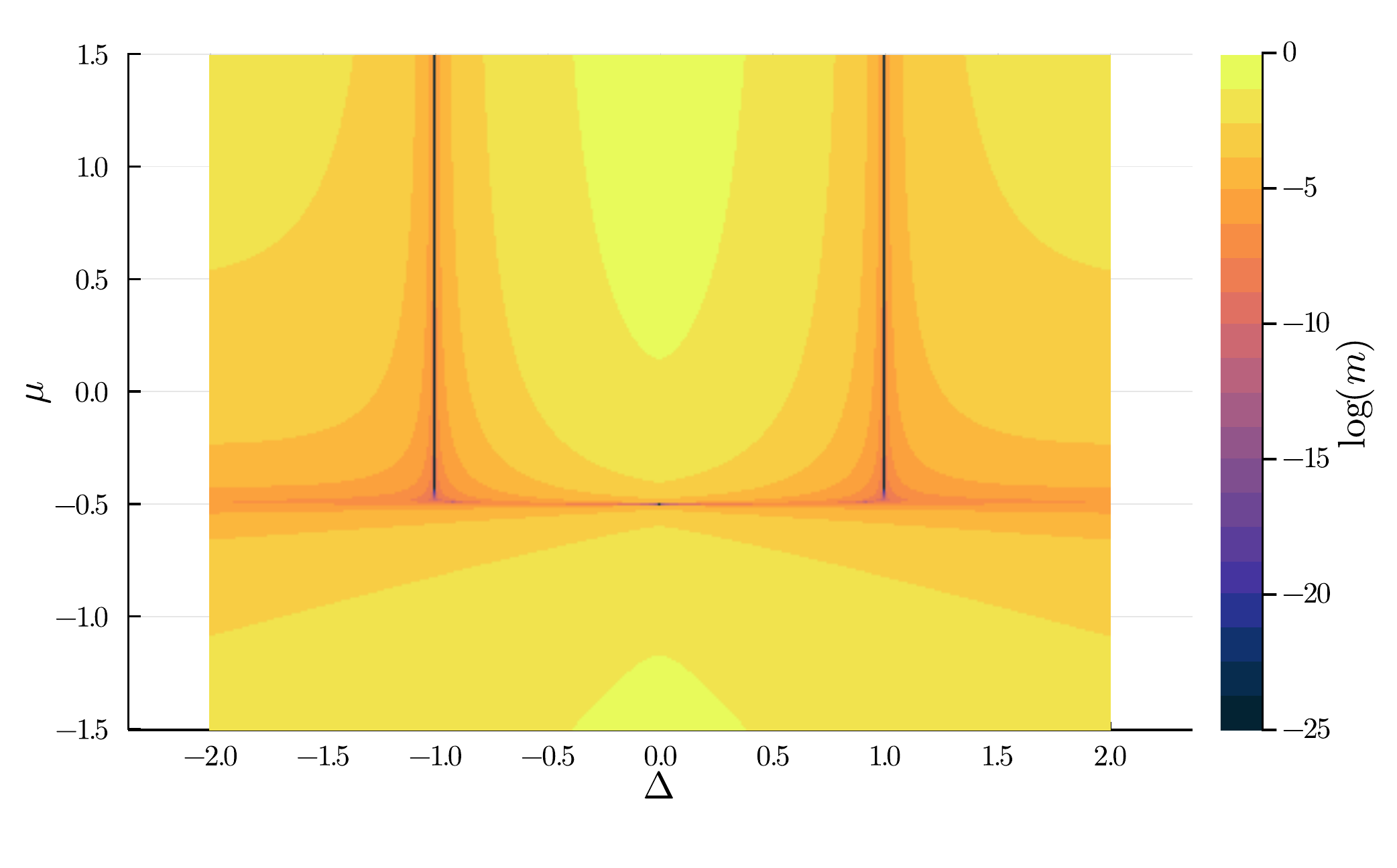}
\caption{Energy gap $m$ phase diagram for the globally coupled fermion model with $N=72$, with logarithmic color scale taken base $e$.}
\label{fig:Ggap}
\end{figure}

\begin{figure}[tb]
\centering
  \includegraphics[width=0.95\linewidth]{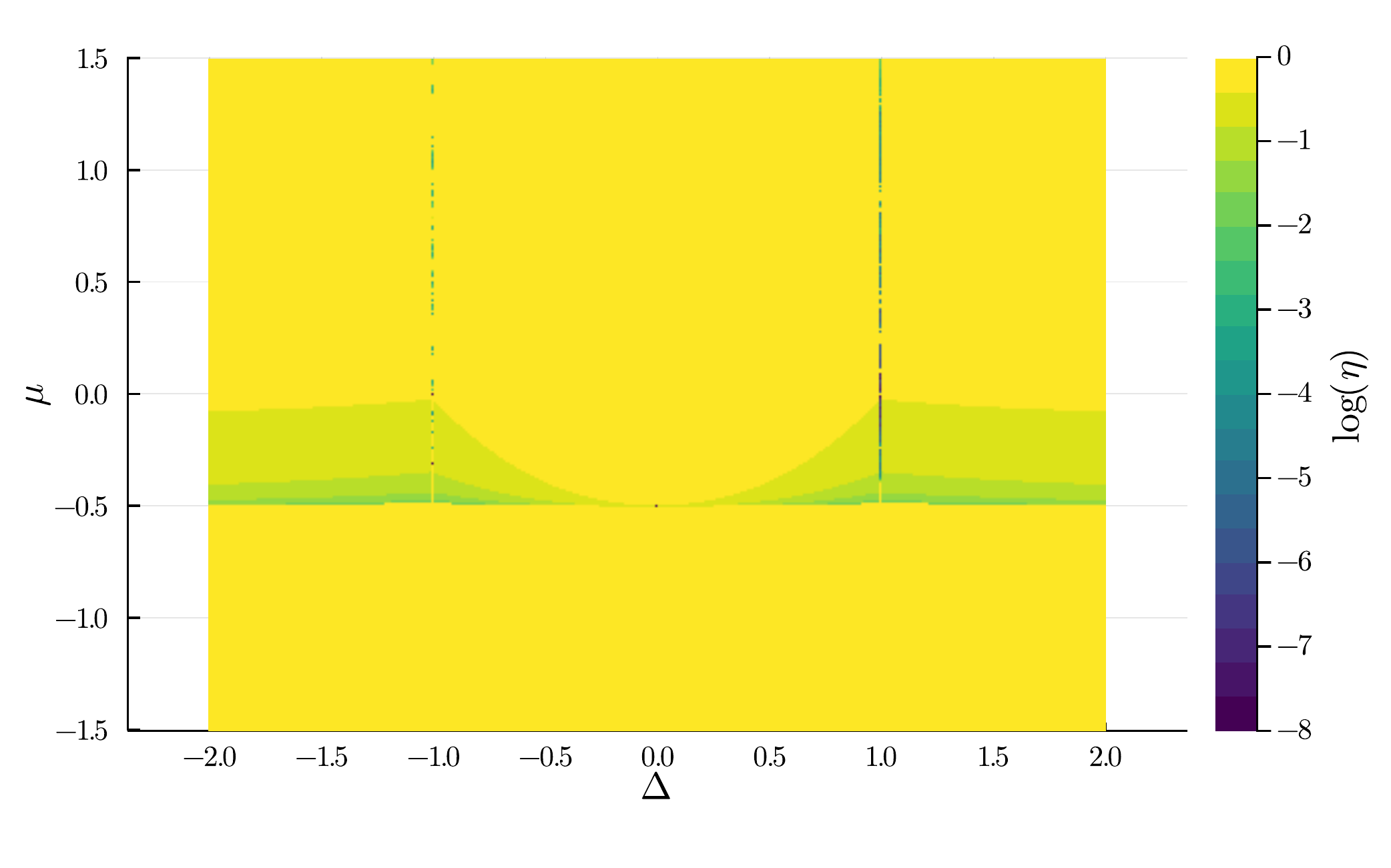}
\caption{Ground state overlap $\eta$ when adding a single site for globally coupled fermion model with $N=72$, with logarithmic color scale taken base $e$.}
\label{fig:Gover}
\end{figure}

In Fig. \ref{fig:Ggap} the phase diagram of the energy gap for different values of $\mu$ and $\Delta$ appear to show several gapped phases with phase transitions on the lines $\mu = -1/2$, and $\Delta = \pm 1, \mu > -1/2$ where the gap closes with the system size. We also see these same regions for the phase diagrams of the ground state overlap in Fig. \ref{fig:Gover}. While this phase diagram is still symmetric for positive and negative $\Delta$, there is a clear asymmetry present in positive and negative $\mu$ which was not present in our previous nearest-neighbor coupled models. 

We plot the scaling of the energy gap with system size for several representative points around $\mu = -0.5, \Delta =1$ in Fig. \ref{fig:gapNG}, where we see the asymptotically constant gap for the extensive phases, a polynomially closing gap for models on the line $\mu = -1/2$, and an exponentially closing gap along the lines $\Delta = \pm 1$ with $\mu > -1/2$.

While the phases with $|\Delta| > 1$ and $\mu > -1/2$ appear to be separate, they are actually connected for finite system size by traversing around the central phase at large $\mu$. For finite number of fermions $N$, the gap closes exactly for $\Delta =0$ and $\mu = -1/2$ or $\mu = (N-1)/2$, so the phase transition lines at $\Delta = \pm 1$ must eventually join when $\mu$ is positive and large enough. However, in the thermodynamic limit the central phase should extent to infinity in the thermodynamic limit, separating the two side phases.

The phase diagram of the ground state overlap for $N=72$ is plotted in Fig. \ref{fig:Gover}, which shows large regions of large overlap, with the smallest overlap occurring close to the line $\mu = -0.5$, but with $\mu > -0.5$, as well as the lines with $\Delta = \pm 1$ where the gap closes exponentially. The finite size scaling of these points in the phase diagram is also shown in Fig. \ref{fig:gapNG}, which shows constant scaling for the selected points in the phase diagram. Notably though, there is a finite system size when the large gap of points at $\mu = -0.49$ transitions from a large value to a smaller value, but seems to stay asymptotically constant. Since all these points are in the vicinity of $\Delta = 1, \mu = -1/2$, it seems that the precise boundaries between these behaviors depend on system size, and all have large overlap for small system sizes, and it is not until larger system size that the points with $\mu > -0.5$ transition to the smaller ground state overlap seen in the phase diagram in Fig. \ref{fig:Gover}. Despite this, it seems these points still have an overlap of near $0.2$ which stays constant with system size. If this is the case, then despite the transition to a smaller gap, the overall complexity is only changed by some constant factor since the overlap still seems to have constant asymptotic scaling.

\section{Recursive state preparation}
\label{sec:recursive}
In the previous sections we have looked at preparing a state by adding one site at a time. This can equivalently be thought of as a process where one first prepares two disjoint ground states, one of size $n-1$ and the other of size $1$, and then ``joins'' them to form the $n$-site ground state with the coupling between the two sub-chains turned on. Nothing in the algorithm restricts us from choosing a different division. For instance, one can prepare two copies of ground states on $n/2$ sites, and then ``join'' them in the middle to form an $n$-site ground state. The runtime of this procedure will now depend on the overlap between the decoupled half-chain ground states and the full ground state. However, since we still only turn on a coupling term between two sites, it is not clear that this overlap should be worse than the overlap between an $n-1$-site and $n$-site ground state. On the other hand, by applying recursion and running disjoint couplings in parallel, one can get from a ground state on a constant number of sites to the full ground state on $n$ sites using $\log n$ steps instead of $n$ steps.

In the single-site scheme one must make a choice of starting state on the ancilla qubit, which affects the overlap. In the recursive scheme, there is no such chioce to be made. Instead, the complexity of the algorithm is determined solely by the energy gaps of the Hamiltonians and the overlap between the tensor product between the ground state of the full Hamiltonian and the tensore product of the ground states on the half-systems. By considering these inner products we can compare the complexity of this recursive scheme to the single-site method. 

%



\begin{figure}[tb]
\centering
   \includegraphics[width=0.95\linewidth]{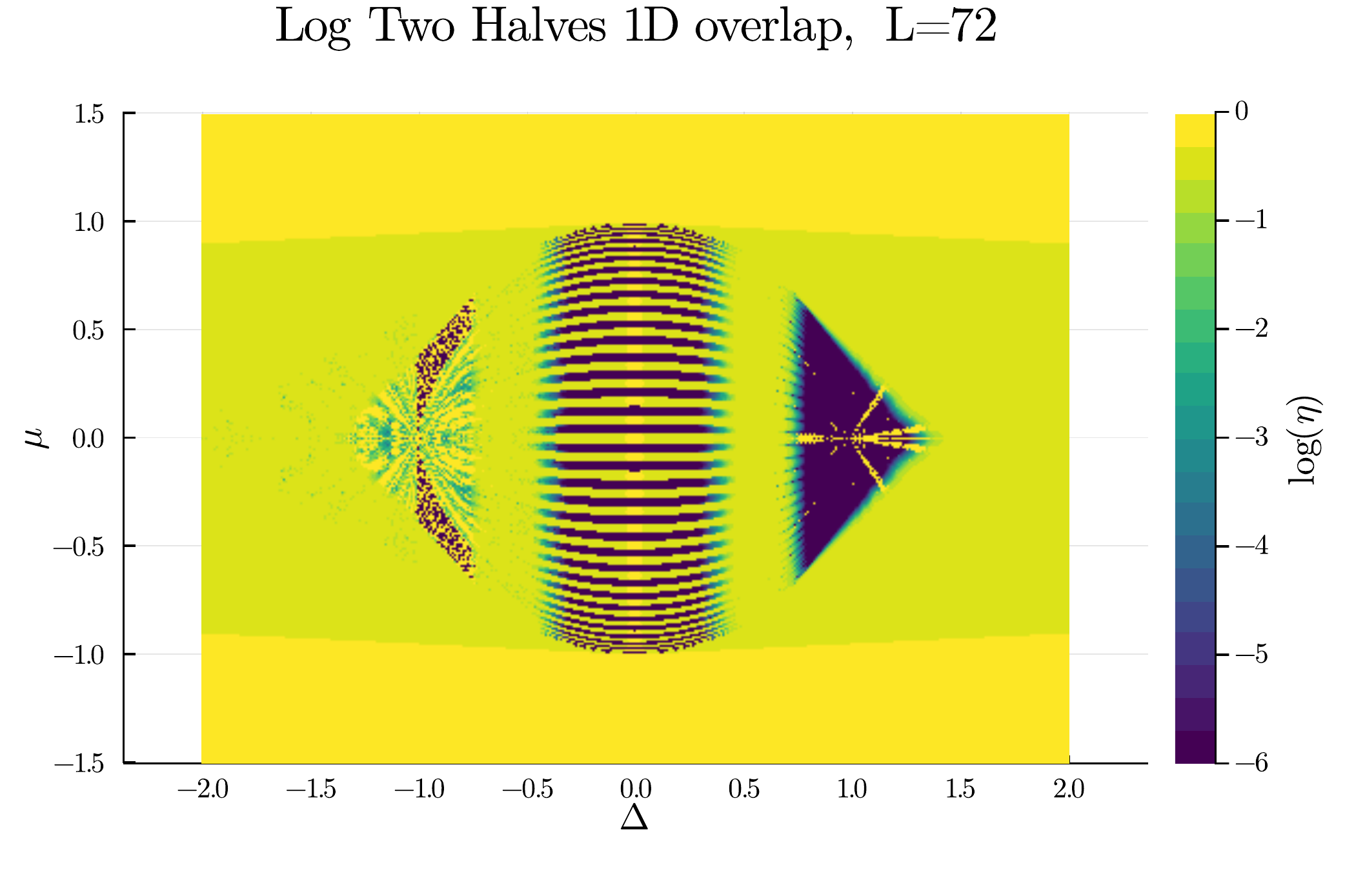}
\caption{Ground state overlap $\eta$ when gluing together two Kitaev chains of equal length, with logarithmic color scale taken base $e$. }
\label{fig:1overhalf}
\end{figure}

\begin{figure}[tb]
\centering
  \includegraphics[width=0.95\linewidth]{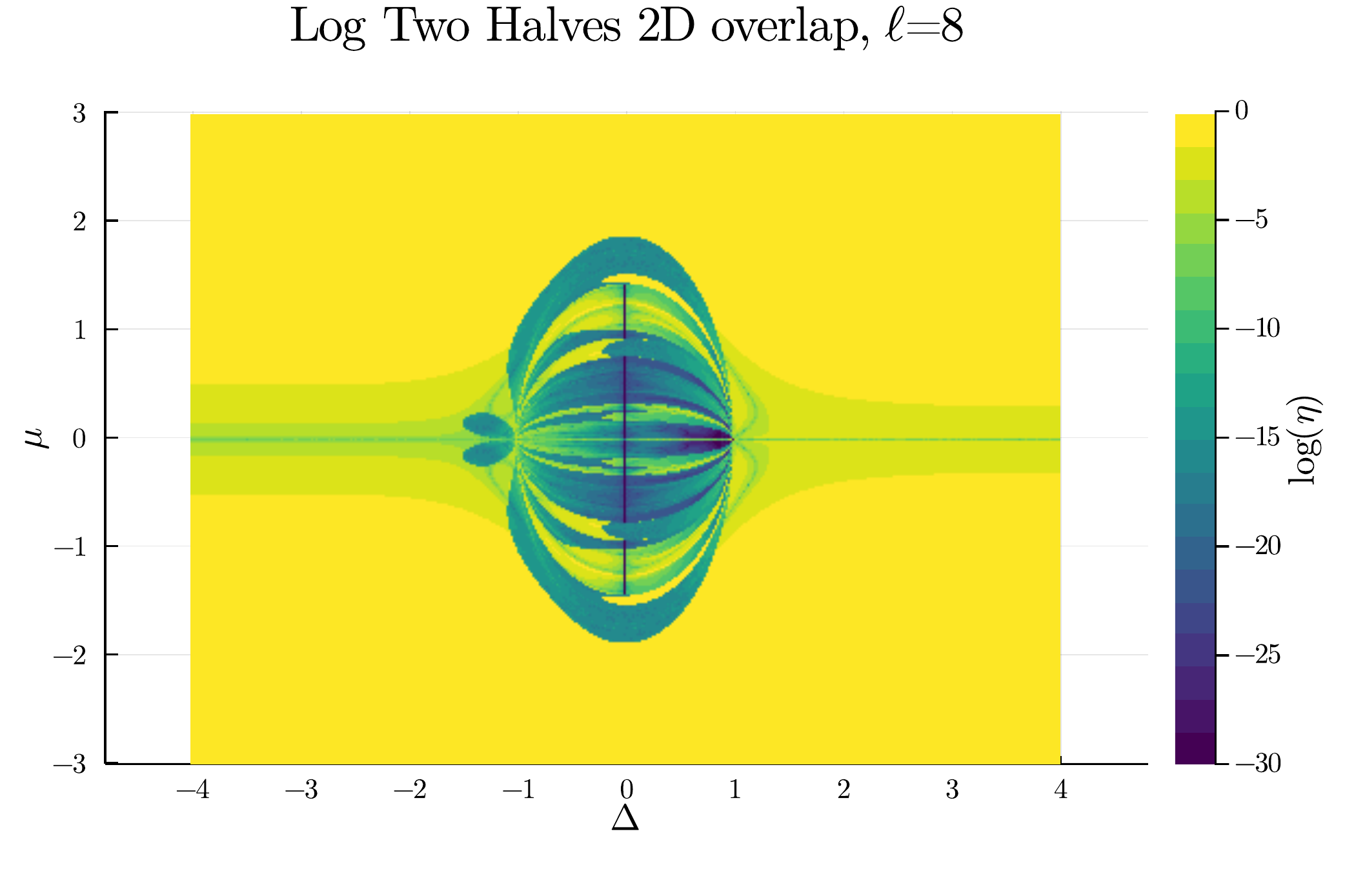}
\caption{Ground state overlap $\eta$ when gluing together two regions of 32 fermions to get a square 2D lattice of 64 fermions, with logarithmic color scale taken base $e$. }
\label{fig:2Doverhalf}
\end{figure}

\begin{figure}[H]
\centering
  \includegraphics[width=0.95\linewidth]{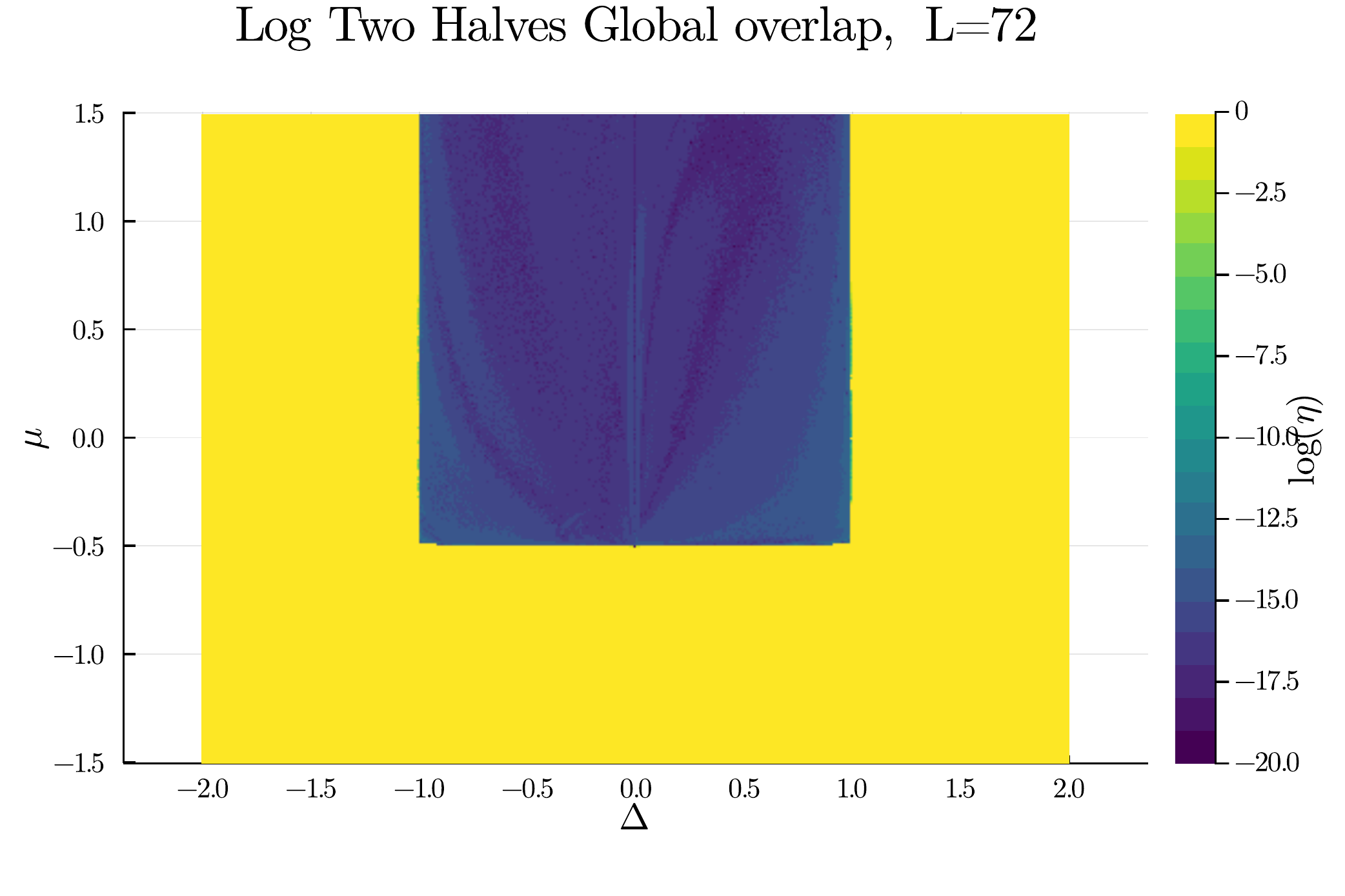}
\caption{Ground state overlap $\eta$ when gluing together two globally coupled clusters of equal size, with ordering all fermions in one cluster are ordered before all fermions in the other, with logarithmic color scale taken base $e$. }
\label{fig:Goverhalf}
\end{figure}

In Figs. \ref{fig:1overhalf}, \ref{fig:2Doverhalf}, \& \ref{fig:Goverhalf} one sees many examples of substantially reduced state overlaps compared to the site-by-site case, such as the central phase of the globally coupled model in Fig. \ref{fig:Goverhalf}. Nevertheless, in many regions of the phase diagrams the overlaps remain large. As discussed in appendix \ref{app_schmidt}, this implies that in these cases there is not too much entanglement across a central cut and the ground state on half the system size has significant overlap with the largest Schmidt state.

The potential benefit of this recursive construction is that, while the overlaps at each step may be reduced, building up many copies of smaller systems to be joined together may be done in parallel to reduce total runtime. 

\section{Conclusion}
\label{sec:conclusion}

In this work we have studied the performance of site-by-site ground state preparation on free-fermion systems. Using computationally efficient formulas, we can access large system sizes and explore the phase space thoroughly at system sizes of up to $N=72$. The quantities of interest are two system parameters, the energy gap and the overlap between subsequent ground states, which together determine the gate complexity of the algorithm. We find that the state overlaps correlate well with the energy gap of the system, sharing many features in their respective phase diagrams. The state overlaps are substantial for large regions of phase space and it is mainly the closing energy gap of critical systems or topological edge modes that impedes the performance of this algorithm. We find that even in some regions where the energy gap closes with system size, such as the $\mu = \pm 1$ critical lines in the Kitaev chain, the ground state overlaps can still display asymptotically constant scaling.

It is interesting to know how well these results extend to broader classes of interacting systems. While it is possible that a favorable scaling of the overlap with system size depends on the underlying Gaussian structure, this relation is not apparent in our numerical results. Indeed, as long as the system is gapped, any perturbative interaction of bounded norm would change the state overlap by a term quadratic in the perturbation strength. In fact, the features present in the gap and overlap phase diagrams can be understood qualitatively via general properties of geometrically local many-body systems such as the correlation length and area laws. Based on this, we expect the state overlaps to behave similarly beyond the free fermion setting as well.

\noindent
\textbf{Acknowledgements:} We thank Yannick Meurice and Ning Bao for useful discussions. This work was supported in part by the U.S. Department of Energy (DOE) under Award Number DE-SC0019139.

\bibliography{refs}
\bibliographystyle{unsrt}

\clearpage

\appendix

\section{Overlap and the Schmidt spectrum}
\label{app_schmidt}
The ground state of the full lattice is in general entangled across any cut. Crucial to the performance of the site-by-site algorithm is the overlap between the full (entangled) ground state and a state that decomposes as a product across some cut. In the special case where we add one site at a time, the initial state is a product on the $(N-1)$-site lattice and the last site. 

We establish bounds on the largest possible overlap between two states where one is a product across a given cut that partitions the system into two parts that we call $L$ and $R$. Consider a Schmidt decomposition of the state, which expresses the entangled state as a positive linear combination of product states,
\begin{equation}
    \ket{\psi} = \suml{i=1}{r}{\sqrt{\lambda_i}\ket{\psi^L_i}\ket{\psi^R_i}}\, ,
\end{equation}
where $r$ denotes the Schmidt rank, and the coefficients $\{\lambda_1,\ldots,\lambda_r\}$ are in descending order.
Suppose we evaluate the inner product of this state with the product state $\ket{\phi} = \ket{\phi^L}\ket{\phi^R}$. Let $a_i = \braket{\phi^L}{\psi^L_i}$ and $b_i = \braket{\phi^R}{\psi^R_i}$, which satisfy $\sum_i|a_i|^2 \leq 1$ and $\sum_i|b_i|^2 \leq 1$. Then, the overlap between the two states is
\begin{equation}
    \left|\braket{\phi_0}{\psi}\right|^2 = \left|\suml{i=1}{r}{\sqrt{\lambda_i}a_i b_i}\right|^2 \le \suml{i=1}{r}{\lambda_i |a_i|^2}\le \lambda_{1}\, ,
\end{equation}
where we first used Cauchy-Schwarz inequality, followed by the fact that $\lambda_1$ is the largest Schmidt coefficient. Therefore, the squared overlap is bounded above by the largest Schmidt coefficient.

A product state with overlap $\lambda_1$ always exists, namely the maximal weight Schmidt state $\ket{\psi^L_1}\ket{\psi^R_1}$, which has $a_1 = b_1 = 1$. The maximal Schmidt coefficient $\lambda_1$ is also always at least $1/r$, with the minimal value obtained for a flat spectrum.  
The Schmidt decomposition is of maximal rank if it is equal to the Hilbert space dimension of the smaller subsystem, i.e. $r = \min (d_L, d_R)$ for $d_L$ and $d_R$ the Hilbert space dimensions of the left and right subsystems.

The largest Schmidt coefficient is also related to the entanglement entropy $S$ as follows: 
\begin{equation}
    e^{-S} = \prodl{i=1}{r}{e^{\lambda_i\ln(\lambda_i)}} = \prodl{i=1}{r}{\lambda_i^{\lambda_i}} \le \prodl{i=1}{r}{\lambda_1^{\lambda_i}} = \lambda_1\, .
\end{equation}
Thus, the entanglement entropy of a state across a bipartition gives a lower bound on the maximal overlap possible with a product state across the same bipartition which is tighter than the bounds given by the Hilbert space dimensions or the Schmidt rank.

In fact, in terms of Renyi entropies $S_k$, the Schmidt rank is equal to $e^{S_0}$, while the largest Schmidt coefficient is equal to $e^{-S_\infty}$.
Each successive Renyi entropy is bounded by the preceding one, $S_k \geq S_{k+1}$. So, the Renyi entropies give a sequence of tighter lower bounds to the maximal Schmidt rank. 

\begin{equation}
    \frac{1}{r} = e^{-S_0} \leq e^{-S_k} \leq e^{-S_\infty} = \lambda_1 
\end{equation}

For the case where one subsystem is a single qubit, an entangled state will have Schmidt rank 2, so there must always exist a product state with overlap at least 1/2, with higher overlap product states existing when the qubit is not maximally entangled.

In practice, the state of the larger subsystem will the ground state of the previous target Hamiltonian, while the smaller subsystem is introduced in a state we are free to choose. 

Let us write the ground state of the system of size $N$ as a Schmidt decomposition of product states on the first $N-k$ sites and the last $k$ sites, with Schmidt rank $r$,

\begin{equation}
    \ket{\psi^N_0} = \suml{i=1}{r}{\sqrt{\lambda_i}\ket{\psi^{N-k}_i}\ket{\psi^k_i}}
\end{equation}

To achieve high overlap it is necessary for the ground state of the previous Hamiltonian, $\ket{\psi^{N-k}_0}$, to have large support over the subspace spanned by the $r$ Schmidt states $\ket{\psi^{N-k}_i}$. Any support outside of this subspace will be a loss in potential overlap, and support on the largest weight Schmidt states allows for the highest potential overlap. Let us also choose $\ket{\psi^{N-k}_{r +1}}$ so that $\ket{\psi^{N-k}_0}$ has support entirely on this subspace of Schmidt vectors extended by one additional state orthogonal to the $r$-dimensional subspace spanned by the Schmidt vectors. Such a choice always exists and can be constructed by, e.g., a Gram-Schmidt process. 

\begin{equation}
    \ket{\psi^{N-k}_0} = \suml{i=1}{r + 1}{a_i\ket{\psi^{N-k}_i}}
\end{equation}

If we prepare the ground state of the $N-k$ site system with some state on the smaller system $\ket{\phi^k} = \sum_i b_i \ket{\psi^k_i}$, the overlap with the $N$ site ground state is upper bounded by

\begin{equation}
    \left |\braket{\psi^N_0}{\psi^{N-k}_0}\ket{\phi^k}\right|^2 = \left|\suml{i=1}{r}{a_i b_i \sqrt{\lambda_i}}\right|^2
    \leq \suml{i=1}{r} \lambda_i |a_i|^2  .
\end{equation}

This upper bound from the Cauchy-Schwarz inequality is achievable by setting $b_i = a^*_i\sqrt{\lambda_i} /\sqrt{\sum_i |a_i|^2 \lambda_i} $.
This quantity is equivalent to the fidelity between the ground state of the $(N-k)$-site system and the reduced density matrix $\sigma$ of the $N$-site ground state on the first $N-k$ sites:

\begin{equation}
    \mathcal{F}(\ket{\psi^{N-k}_0},\sigma) = \bra{\psi^{N-k}_0}\sigma \ket{\psi^{N-k}_0} = \suml{i=1}{r} \lambda_i |a_i|^2 .
\end{equation}

Therefore, while the largest possible overlap is the largest Schmidt weight $\lambda_1$, the best achievable overlap with $\ket{\psi_0^{N-k}}\ket{\phi^k}$ over all possible choices of $k$ is the fidelity of the ($N-k$)-site ground state with the reduced density matrix of the $N$-site ground state on the $N-k$ sites. 


Uhlmann's theorem states that the fidelity between two mixed states is the maximal fidelity of any purification. We may always fix the purification of one state, however, because the fidelity is not changed by a unitary transformation acting only on the purifying subsystem. In our case, we may fix the purification of the mixed state to be the ground state of the larger system, and the maximization occurs over the unentangled degrees of freedom added to the already pure state to match the purified Hilbert space dimension. The new degrees of freedom are introduced in an optimal state to reproduce the fidelity of the smaller mixed state. 

\begin{figure}[tb]
    \centering
      \includegraphics[width=0.95\linewidth]{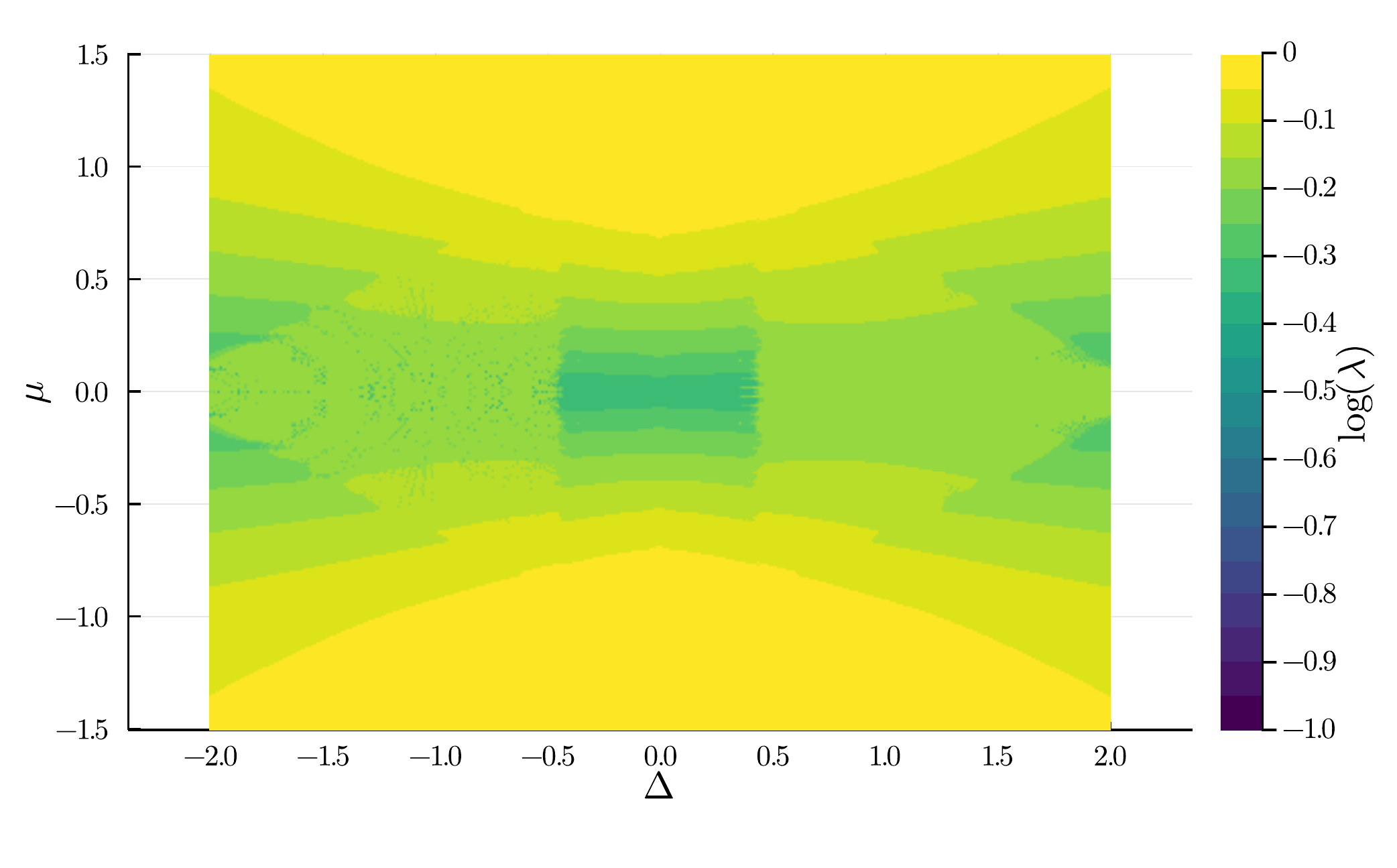}
      \includegraphics[width=0.95\linewidth]{s1ov1d_log.pdf}
    \caption{Top: Largest Schmidt coefficient $\lambda$. Bottom: the actual ground state overlap $\eta$. The former upper bounds the latter. Both are computed for the case of adding a single site for spinless fermion Kitaev chain with open boundary conditions, $N=72$. Color scale is logarithmic taken base $e$.}
    \label{fig:1dover_schmidt}
\end{figure}
\begin{figure}[tb]
    \centering
        \includegraphics[width=0.95\linewidth]{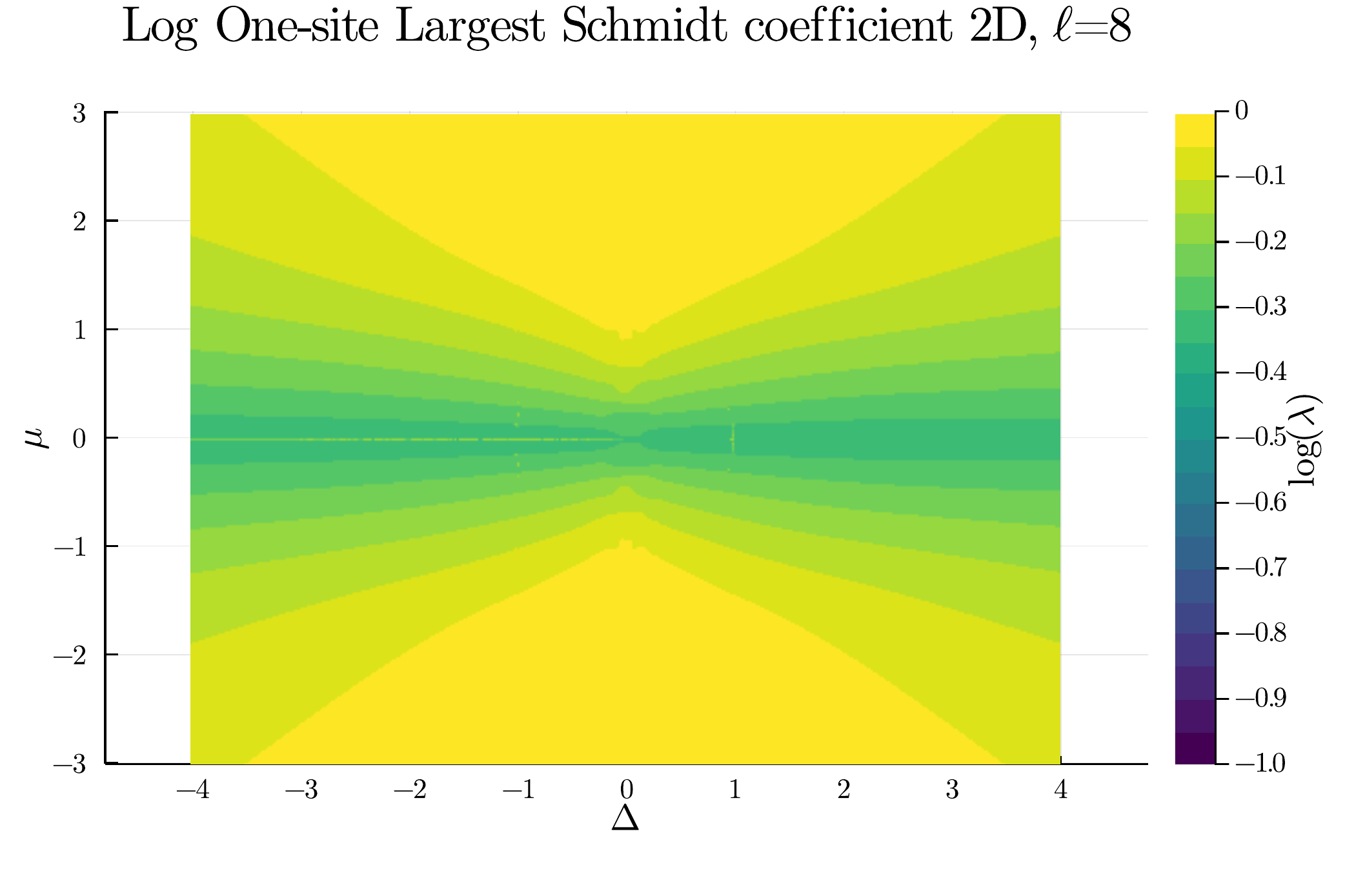}
        \includegraphics[width=0.95\linewidth]{s1ov2d_log.pdf}
    \caption{Largest Schmidt coefficient compared against ground state overlap for the case of adding a single site on the corner to complete square 2D lattice of spinless fermions with side length $\ell = 8$. Color scale is logarithmic taken base $e$.
    \label{fig:2dover_schmidt}}
\end{figure}
\begin{figure}[tb]
    \centering
      \includegraphics[width=0.95\linewidth]{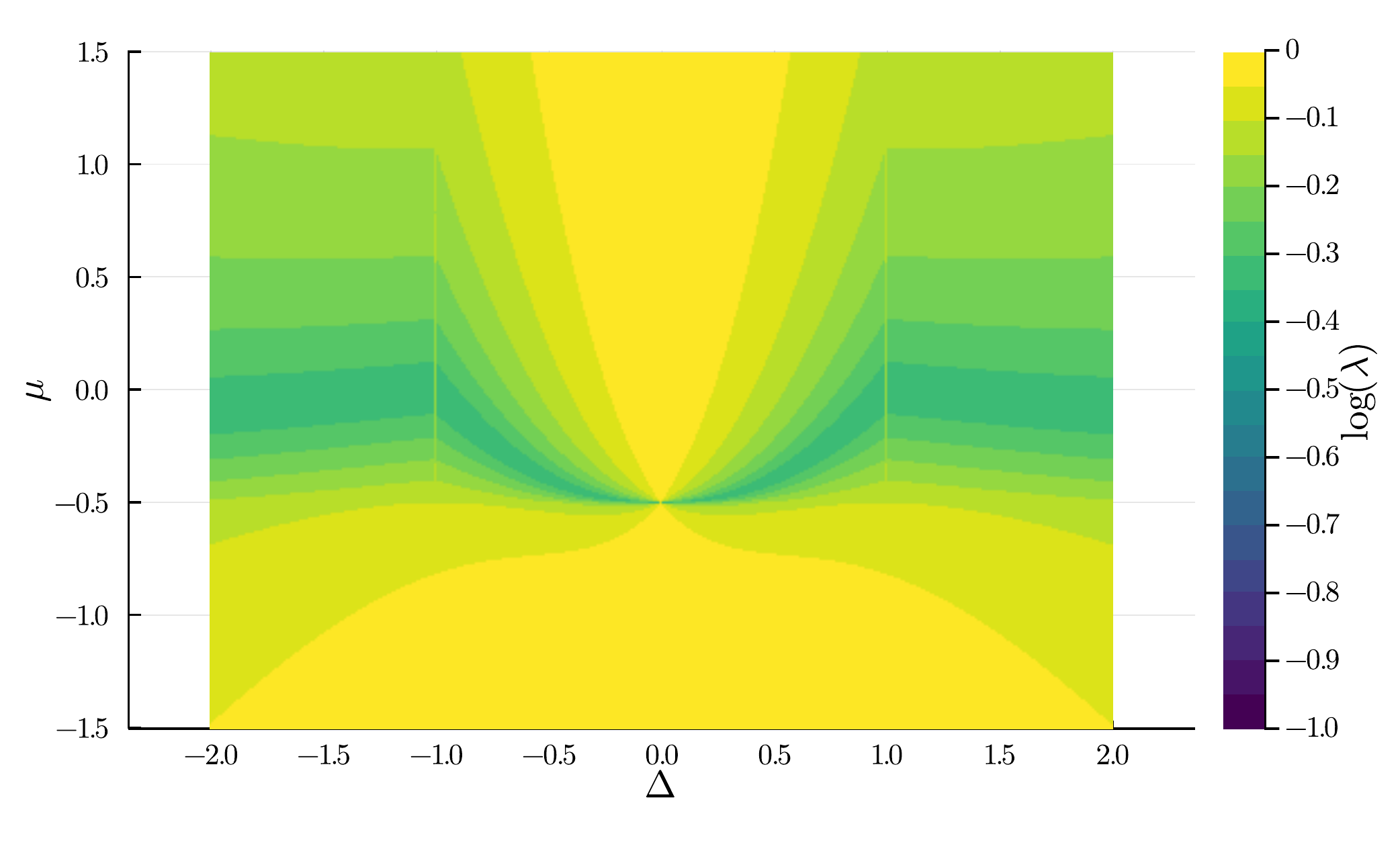}
      \includegraphics[width=0.95\linewidth]{s1ovG_log.pdf}
    \caption{Largest Schmidt coefficient compared against ground state overlap for the case of adding a single site for globally coupled fermion model. Color scale is logarithmic taken base $e$.
    \label{fig:Gover_schmidt}}
\end{figure}
\begin{figure}[tb]
    \centering
       \includegraphics[width=0.95\linewidth]{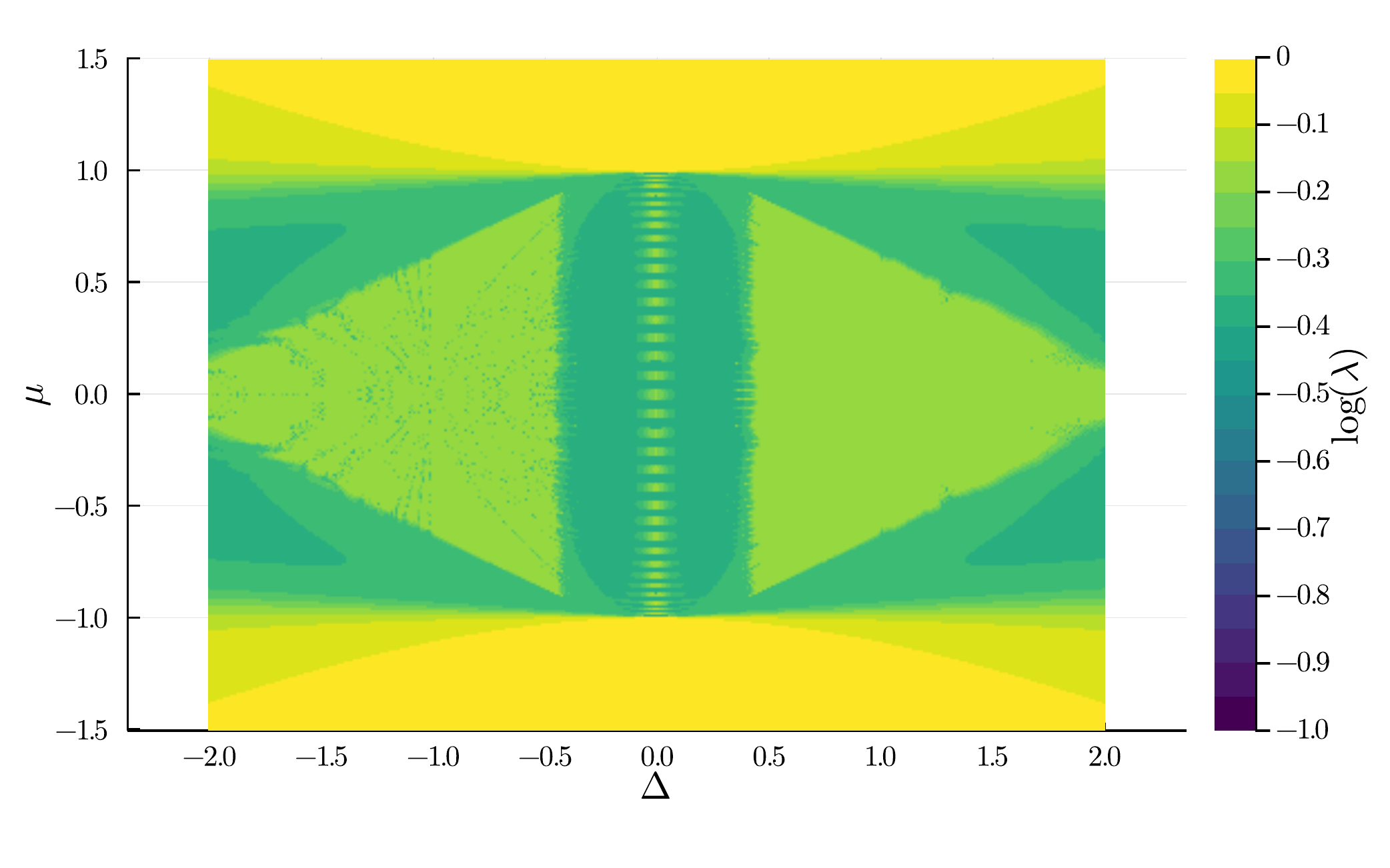}
       \includegraphics[width=0.95\linewidth]{halfov1d_log.pdf}
    \caption{Largest Schmidt coefficient compared against ground state overlap for the case of gluing together two Kitaev chains of equal length, with logarithmic color scale taken base $e$.
    \label{fig:1overhalf_schmidt}}
    \end{figure}
\begin{figure}[tb]
    \centering
      \includegraphics[width=0.95\linewidth]{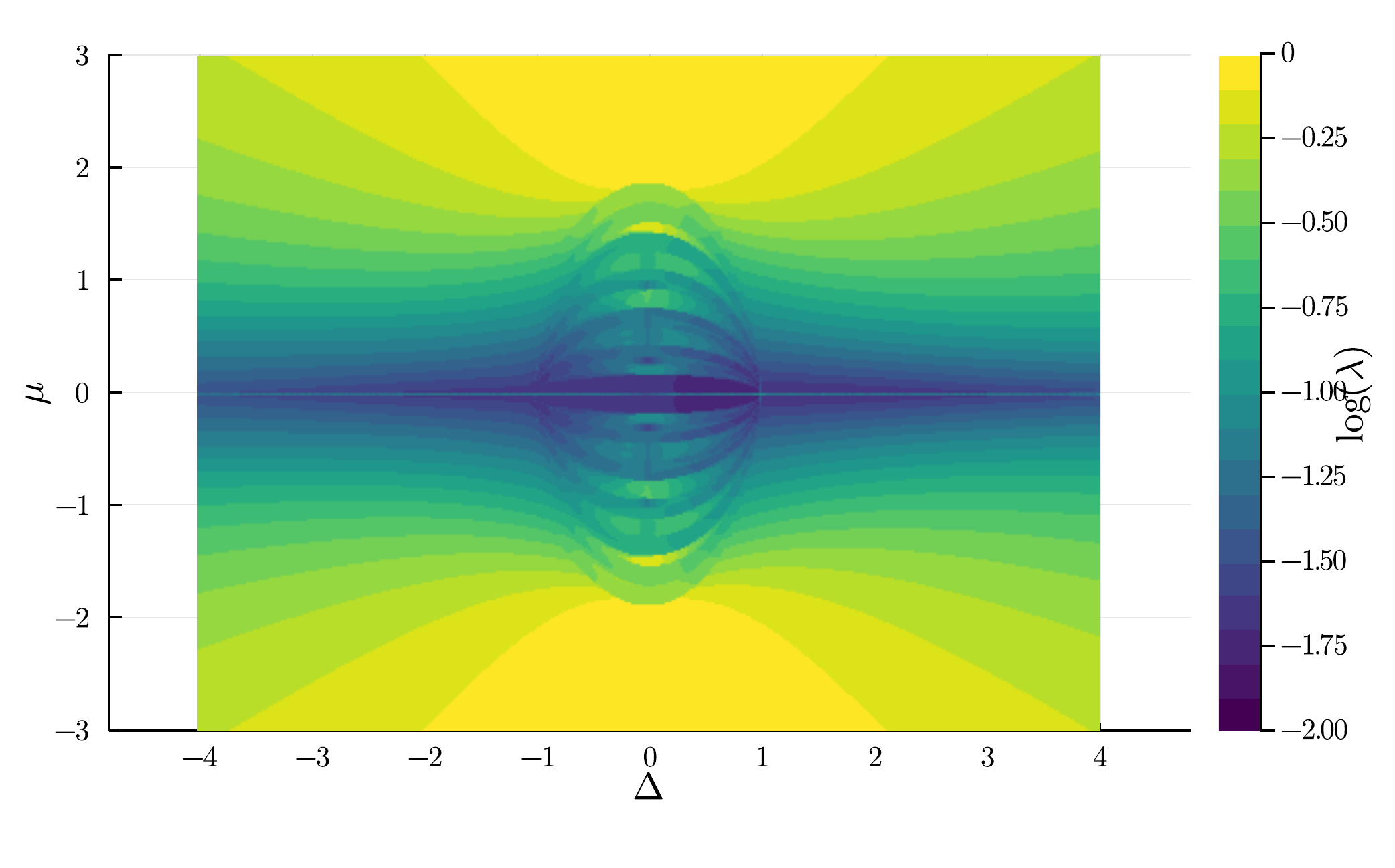}
      \includegraphics[width=0.95\linewidth]{halfov2d_log.pdf}
    \caption{Largest Schmidt coefficient compared against ground state overlap for the case of gluing together two regions of 32 fermions to get a square 2D lattice of 64 fermions, with logarithmic color scale taken base $e$.
    \label{fig:2Doverhalf_schmidt}}
    \end{figure}
\begin{figure}[H]
    \centering
      \includegraphics[width=0.95\linewidth]{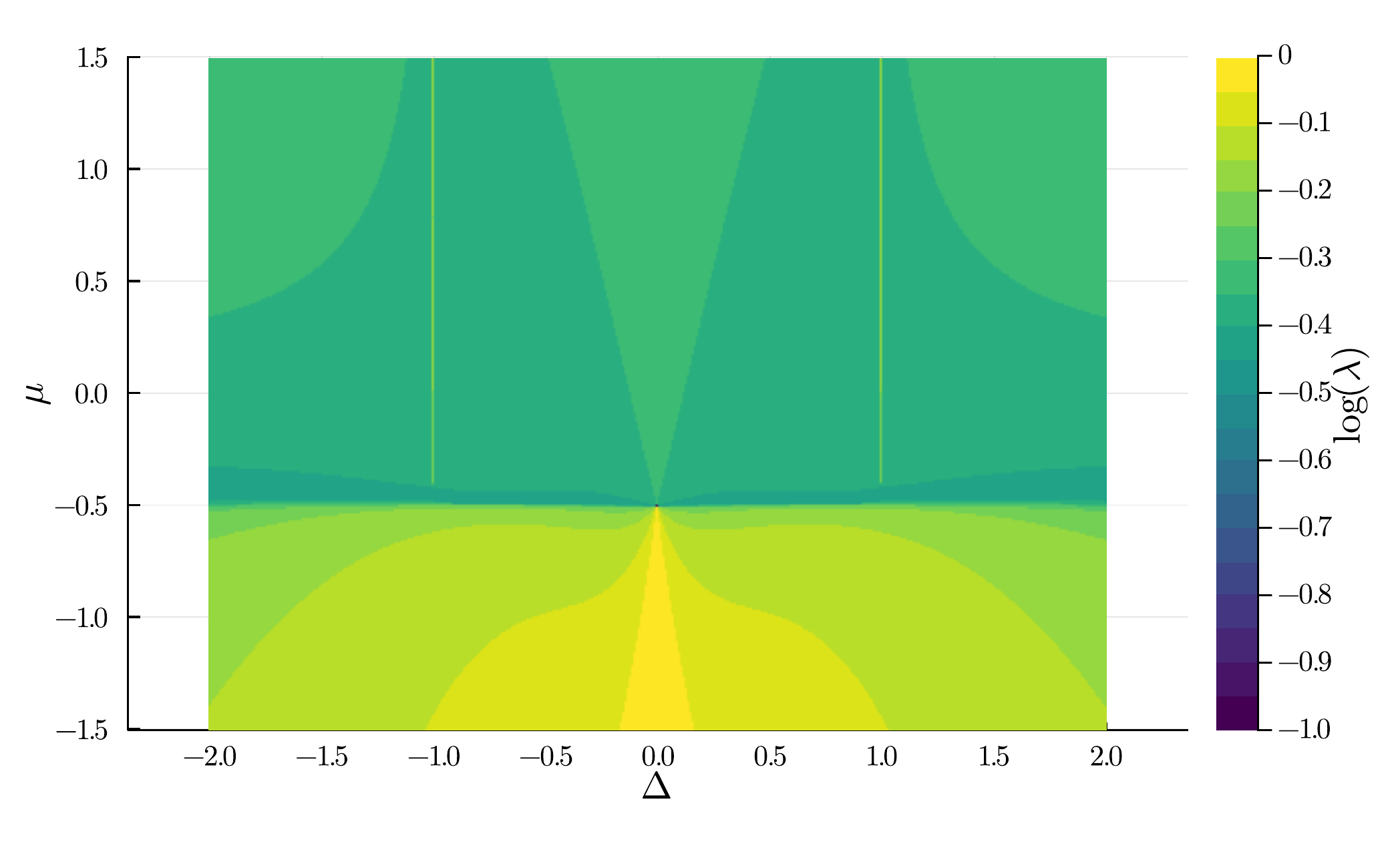}
      \includegraphics[width=0.95\linewidth]{halfovG_log.pdf}
    \caption{Largest Schmidt coefficient compared against ground state overlap for the case of gluing together two globally coupled clusters of equal size, with ordering all fermions in one cluster are ordered before all fermions in the other, with logarithmic color scale taken base $e$.
    \label{fig:Goverhalf_schmidt}}
    \end{figure}

Next, we numerically investigate the tightness of the upper bound on overlap which the Schmidt coefficients provide by comparing these to the overlaps themselves from section \ref{sec:numerics}. As shown in figures \ref{fig:1dover_schmidt}, \ref{fig:2dover_schmidt}, \ref{fig:2dover_schmidt}, \ref{fig:Gover_schmidt}, \ref{fig:1overhalf_schmidt}, \ref{fig:2Doverhalf_schmidt}, and \ref{fig:Goverhalf_schmidt},
this bound is frequently far from tight and provides only a loosely informative guide to the regions in phase space in which the algorithm is efficient.

\end{document}